\documentclass[10pt,pdftex]{article}

\usepackage[english]{babel}
\usepackage{amsmath}
\usepackage{amssymb}
\usepackage[final]{pdfpages}
\usepackage{graphicx}
\usepackage[a4paper, text={16.26cm,23cm},centering]{geometry}

\usepackage{mathtools} 

\usepackage[T1]{fontenc}
\usepackage[utf8]{inputenc}

\makeatletter

\normalsize
\renewcommand\section{\@startsection{section}{1}{\z@}%
                                   {-3.5ex \@plus -1.3ex \@minus -.7ex}%
                                   {2.3ex \@plus.4ex \@minus .4ex}%
                                   {\normalfont\large\bfseries}}
\renewcommand\subsection{\@startsection{subsection}{2}{\z@}%
                                   {-2.3ex\@plus -1ex \@minus -.5ex}%
                                   {1.2ex \@plus .3ex \@minus .3ex}%
                                   {\normalfont\normalsize\bfseries}}
\renewcommand\subsubsection{\@startsection{subsubsection}{3}{\z@}%
                                   {-2.3ex\@plus -1ex \@minus -.5ex}%
                                   {1ex \@plus .2ex \@minus .2ex}%
                                   {\normalfont\normalsize\bfseries}}
\renewcommand\paragraph{\@startsection{paragraph}{4}{\z@}%
                                   {1.75ex \@plus1ex \@minus.2ex}%
                                   {-1em}%
                                   {\normalfont\normalsize\bfseries}}
\renewcommand\subparagraph{\@startsection{subparagraph}{5}{\parindent}%
                                   {1.75ex \@plus1ex \@minus .2ex}%
                                   {-1em}%
                                   {\normalfont\normalsize\bfseries}}

\makeatother

\RequirePackage[sort&compress]{natbib}

\usepackage[pdftitle={General Relativistic Evolution Equations for
  Density Perturbations in Closed, Flat and Open FLRW Universes},
pdfauthor={P. G. Miedema}, pdfsubject={Gauge Invariant Cosmological
  Density Perturbations, Diabatic Perturbations}, pdfstartview={FitH},
pdfkeywords={Cosmological Perturbations, Diabatic Perturbations,
  non-barotropic equations of state}, bookmarks, bookmarksnumbered,
colorlinks=true, linkcolor={blue}, urlcolor={blue},
citecolor={blue}, filecolor=blue]{hyperref}

\newcommand{\nul}{{\scriptscriptstyle(0)}}
\newcommand{\een}{{\scriptscriptstyle(1)}}
\newcommand{\true}{\text{phys}}

\begin{document}

\title{\textbf{\textsf{\noindent General Relativistic Evolution
      Equations for  \protect\\ Density
  Perturbations in Closed, Flat and  \protect\\ Open FLRW Universes}}}

\author{P.\ G.\
  Miedema\footnote{\href{mailto:PG.Miedema@ProtonMail.com}{PG.Miedema@ProtonMail.com}}
  \hfill\\
  Netherlands Defence Academy \\
  Hogeschoollaan 2 \\
  NL-4818CR Breda \\
  The Netherlands}


\date{March 9, 2021}

\maketitle

\begin{abstract}
  It is shown that the decomposition theorems of York, Stewart and
  Walker for symmetric spatial second-rank tensors, such as the
  perturbed metric tensor and perturbed Ricci tensor, and the spatial
  fluid velocity vector imply that, for open, flat or closed
  Friedmann-Lemaître-Robertson-Walker universes, there are exactly
  two, unique, independent gauge-invariant quantities which describe
  the true, physical perturbations to the energy density and particle
  number density.  Using these two new quantities, evolution equations
  for cosmological density perturbations and for entropy
  perturbations, adapted to non-barotropic equations of state for the
  pressure, are derived.  Density perturbations evolve adiabatically
  if and only if the particle number density does not contribute to
  the pressure.  Local density perturbations do not affect the global
  expansion of the universe.  The new perturbation theory has an exact
  non-relativistic limit in a non-static universe.  The gauge problem
  of cosmology has thus been solved.
\end{abstract}

\begin{quote}
\textsc{pacs}: 04.25.Nx; 98.80.-k; 98.80.Jk

\textsc{keywords}: Cosmology; Perturbation theory; mathematical and
relativistic aspects of cosmology
\end{quote}

\newpage
\hrule
\tableofcontents
\bigskip
\hrule

\section{Introduction}
\label{sec:introduction}

The theory related to the linearized Einstein equations and
conservation laws is important in cosmology because it describes the
growth of all kinds of structures in the expanding universe, such as
stars, galaxies and microwave background
fluctuations. \cite{lifshitz1946} and \cite{c15} were the first
researchers to derive a cosmological perturbation theory.  They
encountered the problem that the solutions of the linearized Einstein
equations and conservation laws have no physical significance.  Due to
the linearity of the equations, the solutions can be changed by a
linear coordinate transformation, i.e., a gauge transformation.
Therefore, the solutions are \emph{gauge-dependent}.  This is the
well-known \emph{gauge problem of cosmology} which is up till now not
solved.

\subsection{Gauge Problem of Cosmology}
\label{subsec:gauge-problem-of-cosmolgy}

Consider a closed, flat or open
Friedmann-Lemaître-Robertson-Walker~(\textsc{flrw}) universe filled
with a perfect fluid, with an equation of state for the pressure which
depends, according to thermodynamics, on the particle number
density~$n$ and the energy density~$\varepsilon$. The evolution of the
energy density $\varepsilon_\nul$ and the particle number density
$n_\nul$ of the background universe are governed by the usual Einstein
equations and conservation laws for a \textsc{flrw} universe.  The
general solution of the linearized Einstein equations and conservation
laws contains, among other quantities, $\varepsilon_\een$ and
$n_\een$.  Since the linearized equations are invariant under linear
coordinate transformations
$x^\mu\rightarrow x^{\prime\mu}=x^\mu-\xi^\mu$, one can generate new,
equivalent, solutions $\varepsilon^\prime_\een$ and $n^\prime_\een$,
i.e.,
\begin{equation}\label{eq:gauge-problem}
  \varepsilon^\prime_\een =
     \varepsilon_\een + \xi^0\dot{\varepsilon}_\nul,
  \quad n^\prime_\een = n_\een + \xi^0\dot{n}_\nul,
\end{equation}
where $\xi^0\dot{\varepsilon}_\nul$ and $\xi^0\dot{n}_\nul$ are gauge
modes.  Therefore, $\varepsilon_\een$ and $n_\een$ are
\emph{gauge-dependent} and have, as a consequence, no physical
significance.  In contrast to $\varepsilon_\een$ and $n_\een$ the
real, measurable density perturbations $\varepsilon^\true_\een$ and
$n^\true_\een$ are independent of the choice of a coordinate system,
i.e., are \emph{gauge-invariant}.

The physics of density perturbations is hidden in the general solution
of the linearized Einstein equations and conservation laws.
Therefore, the physical quantities $\varepsilon^\true_\een$ and
$n^\true_\een$ can be expressed as linear combinations, with
time-dependent coefficients, of gauge-dependent solutions of these
equations, such that the gauge modes are eliminated.  Furthermore,
$\varepsilon^\true_\een$ and $n^\true_\een$ have the property that in
the non-relativistic limit the linearized Einstein equations and
conservation laws combined with expressions for
$\varepsilon^\true_\een$ and $n^\true_\een$ reduce, in a
\emph{non-static} flat \textsc{flrw} universe, to the time-independent
Poisson equation with $\varepsilon^\true_\een$ as source term, and the
well-known Einstein relation
$\varepsilon^\true_\een=n^\true_\een mc^2$.  Consequently, the gauge
problem is, in fact, the problem of finding expressions for
$\varepsilon^\true_\een$ and $n^\true_\een$.

\subsection{Previous Attempts to solve the Gauge Problem of Cosmology}

\cite{c13} was the first to demonstrate that using gauge-invariant
quantities, one can recast the linearized Einstein equations and
conservation laws into a new set of evolution equations that are free
of non-physical gauge modes.  Bardeen's definition of a
gauge-invariant density perturbation is such that it becomes equal to
the gauge-dependent density perturbation in the limit of small
scales. The underlying assumption is that on small scales the gauge
modes vanish so that gauge-dependent quantities become
gauge-independent.  As will be shown in Section~\ref{sec:newt-limit},
the physical quantities~$\varepsilon^\true_\een$ and~$n^\true_\een$ do
not become equal to the gauge-dependent quantities~$\varepsilon_\een$
and~$n_\een$ in the non-relativistic limit.  Furthermore, the gauge
modes $\hat{\varepsilon}_\een$ and $\hat{n}_\een$ never become zero,
since the universe is not static in the non-relativistic limit.
Finally, the relativistic space-time gauge transformations reduce in
the non-relativistic limit to Newtonian gauge transformations with
space and time decoupled.  Consequently, the quantities defined by
Bardeen are not equal to the real physical energy density
perturbation.

The article of Bardeen has inspired the pioneering works of
\cite{Ellis1,Ellis2,ellis-1998}, \cite{mfb1992,Mukhanov-2005} and
\cite{Bruni:1992dg}.  These researchers proposed alternative
perturbation theories using gauge-invariant quantities which differ
from the ones used by Bardeen.  The theory of \cite{mfb1992} yields
the Poisson equation of the Newtonian Theory of Gravity only if the
Hubble function, i.e., the expansion scalar, vanishes. This, however,
violates the background Friedmann equation, as will be shown in
Section~\ref{sec:newt-limit}. In fact, the background Friedmann
equation is needed to derive the Poisson equation from the linearized
Friedmann equation in the non-relativistic limit.  Consequently, the
gauge-invariant quantities defined by \cite{mfb1992} are not equal to
the true physical perturbations.

In the approach of \cite{Ellis1,Ellis2,ellis-1998} density
perturbations are defined using gradients. They define a
gauge-invariant and covariant function which ‘closely corresponds to
the intention of the usual gauge-dependent density contrast function'.
However, a perturbation theory based on this definition does not yield
the Poisson equation in the non-relativistic limit.

The above mentioned pioneering treatments caused an abundance of other
important works, e.g., \cite{Malik:2008im,Knobel:2012wa,Peter:2013woa,
  2013GReGr..45.1989M,kodama1984,chung-pei1995}.  The review article
of \cite{Ellis2017} discusses the work of Lifshitz and Khalatnikov,
and also provides an overview of other research on the subject.

From the vast literature on cosmological density perturbations, it
must be concluded that there is still no agreement on which
gauge-invariant quantities are the true energy density perturbation
and particle number density perturbation.  None of the cosmological
perturbation theories in the literature has a correct non-relativistic
limit as explained in Section~\ref{sec:newt-limit}, so that there is
still no solution to the gauge problem of cosmology.

\subsection{Solution of the Gauge Problem of Cosmology}
\label{sec:solution-of-the-gauge-problem}

As said before, the true, physical, density perturbations
$\varepsilon_\een^\true$ and $n_\een^\true$ are hidden in the general
solution of the linearized Einstein equations and conservation laws
and it has proved difficult to determine in advance which linear
combination of gauge-dependent quantities are equal to these true
density perturbations.

To solve this problem, the decomposition theorems of \cite{York1974},
\cite{SteWa} and \cite{Stewart} for symmetric three-tensors are used.
In the literature, these decomposition theorems are only applied to
the perturbed metric three-tensor. This, however, is not sufficient.
The decomposition of the perturbed Ricci three-tensor must also be
taken into account.  Only then the linearized momentum constraint
equations are consistent with the decomposition of the fluid
three-velocity into a rotational part and a divergence part.  This
makes it possible to decompose the linearized equations into three
independent systems which govern the evolution of scalar
perturbations, vector perturbations and tensor perturbations. Only the
part of the perturbed Ricci three-scalar which is associated with
scalar perturbations is non-zero, so that only scalar perturbations
are coupled to $\varepsilon_\een$ and $n_\een$. This fact, which is
explained in detail in Section~\ref{sec:decomp-h-u}, is essential
since for scalar perturbations the three linearized momentum
constraint equations can now be rewritten as one first-order ordinary
differential equation for the perturbed Ricci three-scalar.  This, in
turn, implies that, for scalar perturbations, one exclusively needs
the constraint equations and conservation laws. The fact that these
equations are the perturbed counterparts of the background equations
is crucial.  It is well-known that the evolution of the unperturbed
universe is determined by three independent scalars, namely the energy
density, the particle number density and the expansion scalar, see
Section~\ref{sec:back-eq}.  It has now been achieved that these three
scalars occur also in the linearized equations in
Section~\ref{sec:scalar-pert}, this time as perturbations.  In
Section~\ref{sec:unique} it is shown that this substantially reduces
the number of gauge-invariant quantities which can be written as
linear combinations of gauge-dependent solutions of the linearized
equations.  As a consequence, there are exactly three non-trivial
gauge-invariant quantities, namely, $\varepsilon^{\text{gi}}_\een$,
$n^{\text{gi}}_\een$ and $\theta^{\text{gi}}_\een=0$.  In the
non-relativistic limit, Section~\ref{sec:newt-limit}, it is found that
$\varepsilon^{\text{gi}}_\een\equiv\varepsilon^\true_\een$,
$n^{\text{gi}}_\een\equiv n^\true_\een$ and
$\theta^{\text{gi}}_\een\equiv\theta_\een^\true=0$.  Therefore,
$\varepsilon^{\text{gi}}_\een\equiv\varepsilon^\true_\een$,
$n^{\text{gi}}_\een\equiv n^\true_\een$ and
$\theta^{\text{gi}}_\een\equiv\theta_\een^\true=0$ holds true also in
the General Theory of Relativity.

Finally, evolution equations for $\varepsilon_\een^\true$ and
$n_\een^\true$ are derived using the algorithm given in the appendix,
whereas $\theta_\een^\true=0$ implies that local density perturbations
do not affect the global expansion $\theta_\nul$ of the universe.  In
fact, the cosmological perturbation theory in
Section~\ref{sec:flat-pert} consists of a second-order ordinary
differential equation---with source term entropy perturbations---which
describes the evolution of relative perturbations in the total energy
density, and a first-order ordinary differential equation that
describes the evolution of entropy perturbations.  The coefficients of
the second-order equation depend on the equation of state
$p(n,\varepsilon)$ for the pressure and is, as a consequence,
different from the equations found in the literature.  The first-order
differential equation for the entropy perturbation is a new result.

With the evolution equations for density perturbations given in
Section~\ref{sec:flat-pert}, the gauge problem of cosmology for
closed, flat and open \textsc{flrw} universes has been solved.

\section{Preliminaries}
\label{sec:prelim}

\subsection{Choosing a System of Reference}
\label{subsec:Sys-of-Ref}

In order to derive evolution equations for $\varepsilon_\een^\true$
and $n_\een^\true$ a suitable system of reference must be chosen.  Due
to the general covariance of the Einstein equations and conservation
laws, the General Theory of Relativity is invariant under a general
coordinate transformation $x^\mu\rightarrow x^{\prime\mu}(x^\nu)$, implying
that there are no preferred coordinate systems (see
\cite{weinberg-2008}, Appendix~B).  In particular, the linearized
Einstein equations and conservation laws are invariant under a general
linear coordinate transformation, i.e., a gauge transformation
\begin{equation}
     x^\mu \rightarrow x^{\prime\mu}=x^\mu - \xi^\mu(t,\boldsymbol{x}), \label{func}
\end{equation}
where $\xi^\mu(t,\boldsymbol{x})$ are four arbitrary infinitesimal functions
of time, $x^0=ct$, and space, $\boldsymbol{x}=(x^1,x^2,x^3)$, coordinates, the
so-called gauge functions.  Since there are no preferred systems of
reference and since the solutions $\varepsilon_\een^\true$ and
$n_\een^\true$ of the evolution equations are invariant under the
infinitesimal coordinate transformation~(\ref{func}), i.e., are
gauge-invariant, one may use \emph{any} coordinate system to derive
the evolution equations.

A suitable system of reference can be chosen by the following two
considerations.  Firstly, in order to unambiguously identify the
gauge-invariant quantities $\varepsilon^{\text{gi}}_\een$ and
$n^{\text{gi}}_\een$ found in Section~\ref{sec:unique} as the real
density perturbations $\varepsilon^\true_\een$ and $n^\true_\een$, one
needs the non-relativistic limit.  In the Newtonian Theory of Gravity
space and time are strictly separated, implying that in this theory
\emph{all} coordinate systems are essentially \emph{synchronous}.  In
view of the non-relativistic limit it is, therefore, convenient to use
synchronous coordinates \citep{lifshitz1946,c15,I.12} in the
background as well as in the perturbed universe.  In these coordinates
the metric tensor $g_{\mu\nu}(t,\boldsymbol{x})$ of \textsc{flrw} universes
has the form
\begin{equation}
  \label{eq:metric-flrw}
 g_{00}=1, \quad g_{0i}=0, \quad g_{ij}=-a^2(t)\tilde{g}_{ij}(\boldsymbol{x}), 
\end{equation}
where $a(t)$ is the scale factor of the universe, $g_{00}=1$ indicates
that coordinate time is equal to proper time, $g_{0i}=0$ is the global
synchronicity condition (see \cite{I.12}, \S~84) and
$\tilde{g}_{ij}$\footnote{Quantities with a tilde, i.e., $\tilde{q}$,
  belong to the three-dimensional maximally symmetric subspaces of
  constant time.}  is the metric tensor of the three-dimensional
maximally symmetric sub-spaces of constant time.  From the Killing
equations $\xi_{\mu;\nu}+\xi_{\nu;\mu}=0$ and~(\ref{eq:metric-flrw})
it follows that the functions $\xi^{\mu}(t,\boldsymbol{x})$
in~(\ref{func}) become
\begin{equation}
  \label{eq:synchronous}
\xi^0=\psi(\boldsymbol{x}), \quad
     \xi^i=\tilde{g}^{ik}(\boldsymbol{x})\dfrac{\partial\psi(\boldsymbol{x})}{\partial x^k}
       \int^{ct}\!\!\! \frac{{\text{d}}\tau}{a^2(\tau)} + \chi^i(\boldsymbol{x}),
\end{equation}
if only transformations between synchronous coordinates are allowed.
The four functions $\psi(\boldsymbol{x})$ and $\chi^i(\boldsymbol{x})$ cannot be fixed
since the four coordinate conditions $g_{00}=1$ and $g_{0i}=0$ have
already exhausted all four degrees of freedom, see \cite{c8},
Section~7.4 on coordinate conditions.

Secondly, a property of synchronous systems of reference is that the
space-space components of the four-dimensional Ricci curvature tensor
$R_{\mu\nu}$ are split into two parts such that one part contains
exclusively all time-derivatives of the space-space components of the
metric tensor $g_{ij}$ and the second part $R_{ij}$ is the Ricci
curvature tensor of the three-dimensional sub-spaces.  In fact,
$R_{ij}$ is the three-dimensional Ricci tensor which is expressed in
terms of $g_{ij}$ in the same way as $R_{\mu\nu}$ is expressed in
terms of $g_{\mu\nu}$. All operations of raising indices and of
covariant differentiation are carried out with the three-dimensional
metric $g_{ij}$, \cite{I.12}, \S~97.  These properties make it
possible to apply the decomposition theorems of \cite{York1974},
\cite{SteWa} and \cite{Stewart} to the perturbed metric three-tensor
and perturbed Ricci three-tensor.

In conclusion, a synchronous system of reference is the most
appropriate coordinate system to derive the evolution equations for
the density perturbations $\varepsilon^\true_\een$ and $n^\true_\een$.

\subsection{Equations of State for the Pressure}
\label{subsec:eq-of-state}

Barotropic equations of state $p=p(\varepsilon)$ do not take into
account the particle number density, so that the influence of entropy
perturbations on the evolution of density perturbations cannot be
studied, see Section~\ref{subsec:adiabatic-pert}.  Since heat exchange
of a perturbation with its environment could be important for
structure formation in the early universe, realistic equations of
state are needed.  From thermodynamics it is known that both the
energy density $\varepsilon$ and the pressure $p$ depend on the
particle number density $n$ and the absolute temperature $T$, i.e.,
\begin{equation}
  \label{eq:es-p-T}
  \varepsilon=\varepsilon(n,T), \quad
      p=p(n,T).
\end{equation}
Since $T$ can, in principle, be eliminated from these equations of
state, a computationally more convenient equation of state for the
pressure is used, namely
\begin{equation}
  \label{eq:equat-of-state-pressure}
    p=p(n,\varepsilon).
\end{equation}
This equation of state for the pressure and its first- and
second-order partial derivatives have been included in the evolution
equations for the density perturbations $\varepsilon_\een^\true$ and
$n^\true_\een$.

\section{Einstein Equations and Conservation laws for Closed, Flat
  and  \protect\\ Open FLRW Universes}
\label{sec:basic-equations}

The background and linearized equations can be calculated from the set
of Einstein equations (97.11)--(97.13) from the textbook of
\cite{I.12} and the conservation laws ${T^{\mu\nu}{}_{;\nu}=0}$.

\subsection{Background Equations}
\label{sec:back-eq}

The complete set of background Einstein equations and conservation
laws for open, flat and closed \textsc{flrw} universes filled with a
perfect fluid with energy-momentum tensor
\begin{equation}
  \label{eq:en-mom-tensor}
  T^{\mu\nu}=(\varepsilon+p)u^\mu u^\nu - p g^{\mu\nu}, \quad 
p=p(n,\varepsilon),
\end{equation}
is, in synchronous coordinates, given by 
\begin{subequations}
\label{subeq:einstein-flrw}
\begin{alignat}{3}
   3H^2 & =\tfrac{1}{2}R_\nul+\kappa\varepsilon_\nul+\Lambda, \quad &
             \kappa& =8\pi G_{\text{N}}/c^4, & \label{FRW3}\\
  \dot{R}_\nul & =-2HR_\nul, & & \label{FRW3a}\\
  \dot{\varepsilon}_\nul & = -3H\varepsilon_\nul(1+w), & 
           w & \coloneqq p_\nul/\varepsilon_\nul, & \label{FRW2} \\
  \vartheta_\nul & =0, & & \label{FRW-theta-0} \\
  \dot{n}_\nul & = -3Hn_\nul. & &  \label{FRW2a}
\end{alignat}
\end{subequations}
The $G_{0i}$ constraint equations and the $G_{ij}$, $i\neq j$,
dynamical equations are identically satisfied. The $G_{ii}$ dynamical
equations are equivalent to the time-derivative of the $G_{00}$
constraint equation, or Friedmann equation~(\ref{FRW3}).  Therefore,
the $G_{ij}$ dynamical equations need not be taken into account.  In
equations~(\ref{subeq:einstein-flrw}) $\Lambda$ is the cosmological
constant, $G_{\text{N}}$ the gravitational constant of the Newtonian
Theory of Gravity and $c$ the speed of light.  An over-dot denotes
differentiation with respect to $ct$ and the sub-index $(0)$ refers to
the background, i.e., unperturbed, quantities. Furthermore,
$H\coloneqq\dot{a}/a$ is the Hubble function which is, for
\textsc{flrw} universes, equal to $H=\tfrac{1}{3}\theta_\nul$, where
$\theta_\nul$ is the background value of the expansion scalar
$\theta\coloneqq u^\mu{}_{;\mu}$ with $u^\mu\coloneqq c^{-1}U^\mu$ the
fluid four-velocity, normalised to unity ($u^\mu u_\mu=1$).  A
semicolon denotes covariant differentiation with respect to the
background metric tensor $g_{\nul\mu\nu}$.  The \emph{spatial} parts
of the background Riemann tensor $R^i_{\nul jkl}$, the Ricci curvature
tensor $R^i_{\nul j}$ and its contraction $R_\nul$ are given~by
\begin{equation}
  \label{eq:glob-curve}
     R^i_{\nul jkl}=\tilde{R}^i{}_{jkl}=K\left(\delta^i{}_k \tilde{g}_{jl}-
         \delta^i{}_l\tilde{g}_{jk}\right), \quad  R^i_{\nul j}=-\dfrac{2K}{a^2}\delta^i{}_j,
      \quad R_\nul=-\dfrac{6K}{a^2}, \quad  K=-1,0,+1,
\end{equation}
where $R_\nul$ is the global spatial curvature.  The quantity
$\vartheta_\nul$ is the three-divergence of the spatial part of the
fluid four-velocity~$u_\nul^\mu$. For an isotropic universe the fluid
four-velocity is $u^\mu_\nul=\delta^\mu{}_0$, implying that
$\vartheta_\nul=0$, so that there is no \emph{local} fluid flow.

From the system~(\ref{subeq:einstein-flrw}) one can conclude that the
evolution of an unperturbed \textsc{flrw} universe is determined by
exactly three independent scalars, namely
\begin{equation}\label{eq:scalars-flrw}
    \varepsilon = T^{\mu\nu} u_\mu u_\nu, \quad
    n = N^\mu u_\mu, \quad
    \theta  = u^\mu{}_{;\mu},
\end{equation}
where $N^\mu\coloneqq nu^\mu$ is the cosmological particle current
four-vector, which satisfies the particle number conservation law
${N^\mu{}_{;\mu}=0}$,~(\ref{FRW2a}), see \cite{weinberg-2008},
Appendix~B\@.  As will become clear in Sections~\ref{sec:unique}
and~\ref{sec:newt-limit}, the quantities~(\ref{eq:scalars-flrw}) and
their perturbed counterparts play a key role in the evolution of
cosmological density perturbations.

\subsection{Linearized Equations}
\label{sec:pert-eq}

The complete set of linearized Einstein equations and conservation
laws for open, flat and closed \textsc{flrw} universes is, in
synchronous coordinates, given by
\begin{subequations}
\label{subeq:basis}
\begin{align}
  &  H\dot{h}^k{}_k +
\tfrac{1}{2}R_\een =
         -\kappa\varepsilon_\een,     \label{basis-1} \\
  & \dot{h}^k{}_{k|i}-\dot{h}^k{}_{i|k} =
         2\kappa(\varepsilon_\nul + p_\nul) u_{\een i}, \label{basis-2} \\
  & \ddot{h}^i{}_j+ 3H\dot{h}^i{}_j +\delta^i{}_j H\dot{h}^k{}_k
      +2R^i_{\een j}= -\kappa\delta^i{}_j(\varepsilon_\een-p_\een),
         \label{basis-3} \\
  & \dot{\varepsilon}_\een + 3H(\varepsilon_\een+p_\een)+
       (\varepsilon_\nul +p_\nul)\theta_\een=0,
               \label{basis-4}  \\
  & \frac{1}{c}\frac{{\text{d}}}{{\text{d}} t}
    \Bigl[(\varepsilon_\nul+p_\nul) u^i_\een\Bigr]-g^{ik}_\nul p_{\een|k}+
    5H(\varepsilon_\nul+p_\nul) u^i_\een=0,
            \label{basis-5} \\
  & \dot{n}_\een+3Hn_\een+n_\nul\theta_\een = 0,  \label{basis-6}
\end{align}
\end{subequations}
where $h_{\mu\nu}\coloneqq -g_{\een\mu\nu}$ and
$h^{\mu\nu}=+g_\een^{\mu\nu}$ with $h_{00}=0$ and $h_{0i}=0$ is the
perturbed metric tensor, $h^i{}_j=g_\nul^{ik}h_{kj}$, and
$g_\nul^{ij}$ is the unperturbed background metric
tensor~(\ref{eq:metric-flrw}) for an open, flat and closed
\textsc{flrw} universe.  Quantities with a sub-index $(1)$ are the
perturbed counterparts of the background quantities with a sub-index
$(0)$.  A vertical bar denotes covariant differentiation with respect
to the background metric tensor $g_{\nul ij}$.  Note that the
$G^0_{\een0}$ and $G^0_{\een i}$ constraint equations~(\ref{basis-1})
and~(\ref{basis-2}) contain at most first-order time-derivatives of
the metric.  They relate the initial conditions $h^i{}_j(t_0,\boldsymbol{x})$,
$\dot{h}^i{}_j(t_0,\boldsymbol{x})$, $\varepsilon_\een(t_0,\boldsymbol{x})$,
$n_\een(t_0,\boldsymbol{x})$ and $u^i_\een(t_0,\boldsymbol{x})$. Once the constraint
equations are satisfied at $t=t_0$, they are satisfied automatically
at all times.  This is an intrinsic property of the Einstein equations
(\cite{c8} and \cite{I.12}, \S~95), which will be used in
Section~\ref{sec:scalar-pert}.

The perturbation to the pressure is given by the perturbed equation of
state for the pressure
\begin{equation}
\label{eq:p1}
p_\een=p_nn_\een+p_\varepsilon\varepsilon_\een, \quad
  p_n\coloneqq\left(\dfrac{\partial p}{\partial n}\right)_{\!\!\varepsilon}, \quad
p_\varepsilon\coloneqq\left(\dfrac{\partial p}{\partial \varepsilon}\right)_{\!\!n},
\end{equation}
where $p_n(n,\varepsilon)$ and $p_\varepsilon(n,\varepsilon)$ are the
partial derivatives of the equation of state $p(n,\varepsilon)$.

Using Lifshitz' formula, see \cite{c15}, equation (I.3) and \cite{c8},
equation (10.9.1),
\begin{equation}
    \Gamma^k_{\een ij}=
   -\tfrac{1}{2} g^{kl}_\nul
     (h_{li|j}+h_{lj|i}-h_{ij|l}), \label{con3pert}
\end{equation}
and the contracted Palatini identities, see~\cite{c15}, equation~(I.5)
and~\cite{c8}, equation (10.9.2),
\begin{equation}
    R_{\een ij} =
\Gamma^k_{\een ij|k}-\Gamma^k_{\een ik|j},
    \label{palatini}
\end{equation}
the perturbed Ricci three-tensor can be written as
\begin{equation}
  \label{eq:Ricci-lower}
  R_{\een ij}=-\tfrac{1}{2} g^{kl}_\nul (h_{li|j|k}+h_{lj|i|k}-h_{ij|l|k}-h_{lk|i|j}).
\end{equation}
Raising the index~$i$, one gets, using also~(\ref{eq:glob-curve}),
\begin{equation}
  \label{eq:ricci-1}
     R^i_{\een j}\coloneqq (g^{ik}R_{kj})_\een=
     g^{ik}_\nul R_{\een kj}+\tfrac{1}{3}R_\nul h^i{}_j=
     -\tfrac{1}{2}g^{il}_\nul(h^k{}_{l|j|k}+h^k{}_{j|l|k}-h^k{}_{k|l|j})+
      \tfrac{1}{2}g^{kl}_\nul h^i{}_{j|k|l} +
      \tfrac{1}{3}R_\nul h^i{}_j.
\end{equation}
Using that $g_\nul^{ij}h^k{}_{i|j|k}=g_\nul^{ij}h^k{}_{i|k|j}$, one
finds for the contraction of the perturbed Ricci three-tensor, i.e.,
the perturbed Ricci three-scalar,
\begin{equation}
    R_\een  \coloneqq R^k_{\een k} =
   g_\nul^{ij} (h^k{}_{k|i|j}-h^k{}_{i|k|j}) +
     \tfrac{1}{3}R_\nul h^k{}_k.   \label{driekrom}
\end{equation}
Expression~(\ref{driekrom}) is the local perturbation to the global
spatial curvature $R_\nul$ due to a local density perturbation.

Finally, $\theta_\een$ is the perturbation to the expansion scalar
$\theta\coloneqq u^\mu{}_{;\mu}$.  Using that
$u^\mu_\nul=\delta^\mu{}_0$, one gets
\begin{equation}
  \theta_\een=
  \vartheta_\een-\tfrac{1}{2}\dot{h}^k{}_k, \quad
  \vartheta_\een\coloneqq u^k_{\een|k}, 
  \label{fes5}
\end{equation}
where $\vartheta_\een$ is the divergence of the spatial part of the
perturbed fluid four-velocity $u^\mu_\een$.

\section{Decomposition of the Spatial Metric
  Tensor, the Spatial Ricci Tensor \protect\\ and  the Spatial Fluid
  Velocity}
\label{sec:decomp-h-u}

\cite{York1974}, \cite{SteWa} and \cite{Stewart} showed that
\emph{any} symmetric second-rank three-tensor, and hence the
perturbation tensor $h_{ij}$ and the perturbed Ricci tensor
$R_{\een ij}$, can uniquely be decomposed into three parts.  For the
perturbed metric tensor, one has
\begin{equation}
  h^i{}_j  =h^i_{\parallel j} + h^i_{\perp j} + h^i_{\ast j}, \qquad
   h^k_{\perp k}=0, \quad  h^k_{\ast k}=0, \quad  h^k_{\ast i|k}=0.    \label{eq:h123}
\end{equation}
The perturbed Ricci tensor, $R_{\een ij}$, being a symmetric
second-rank three-tensor, can in the same way be decomposed
\begin{equation}
 \label{eq:prop-Rij}
  R^i_{\een j}  =R^i_{\een\parallel j} + R^i_{\een\perp j} + R^i_{\een\ast j}, \qquad
R^k_{\een\perp k}=0, \quad  R^k_{\een\ast k}=0, \quad  R^k_{\een\ast i|k}=0.
\end{equation}
Finally, York and Stewart demonstrated that the components
$h^i_{\parallel j}$ can be written in terms of two independent
potentials $\phi(t,\boldsymbol{x})$ and $\zeta(t,\boldsymbol{x})$, namely
\begin{equation}
  h^i_{\parallel j} =
      \frac{2}{c^2}(\phi\delta^i{}_j+\zeta^{|i}{}_{|j}).
        \label{decomp-hij-par}
\end{equation}
As will be shown in the next three subsections, the
decompositions~(\ref{eq:h123}) and~(\ref{eq:prop-Rij}) and the
expression~(\ref{decomp-hij-par}) imply that the decomposition of the
spatial part $\boldsymbol{u}_\een=(u^1_\een,u^2_\een,u^3_\een)$ of the
perturbed fluid four-velocity
\begin{equation}
   \boldsymbol{u}_{\een} = \boldsymbol{u}_{\een\parallel} +
  \boldsymbol{u}_{\een\perp}, \label{eq:decomp-u}
\end{equation}
where the components $\boldsymbol{u}_{\een\parallel}$ and $\boldsymbol{u}_{\een\perp}$
have the properties
\begin{equation} 
  \boldsymbol{\tilde{\nabla}}\cdot\boldsymbol{u}_\een=\boldsymbol{\tilde{\nabla}}\cdot\boldsymbol{u}_{\een\parallel}, \quad
\boldsymbol{\tilde{\nabla}}\times\boldsymbol{u}_\een=\boldsymbol{\tilde{\nabla}}\times\boldsymbol{u}_{\een\perp}, \label{eq:nabla-u}
\end{equation}
is consistent with the momentum constraint equations~(\ref{basis-2}).
In expressions~(\ref{eq:nabla-u}) $\boldsymbol{\tilde{\nabla}}$ is the
generalised vector differential operator, defined by
$\tilde{\nabla}_i v^k\coloneqq v^k{}_{|i}$.

Using the decompositions~(\ref{eq:h123}), (\ref{eq:prop-Rij})
and~(\ref{eq:decomp-u}) the system of equations~(\ref{subeq:basis}) is
split into three independent systems. The solutions of these systems
are referred to as tensor perturbations~$\ast$, vector
perturbations~$\perp$ and scalar perturbations~$\parallel$. These
perturbations are briefly discussed in the following three
subsections.

\subsection{Tensor Perturbations}

Using~(\ref{eq:h123}) and~(\ref{eq:prop-Rij}),
equations~(\ref{basis-1})--(\ref{basis-3}) imply that
$\varepsilon_\een=0$, $p_\een=0$ and $\boldsymbol{u}_\een=0$, so that tensor
perturbations are not coupled to density
perturbations. With~(\ref{fes5}) it follows that
$\theta_\een=0$. This, in turn, implies that equation~(\ref{FRW2a})
and~(\ref{basis-6}) are identical, so that $n_\een=0$. As a
consequence, the system~(\ref{subeq:basis}) reduces for tensor
perturbations to
\begin{equation}
  \label{eq:tensor-perturbations}
  \ddot{h}^i_{\ast j}+3H\dot{h}^i_{\ast j}+2R^i_{\een\ast j}=0.
\end{equation}
Due to the form of this equation, tensor perturbations are usually
referred to as \emph{gravitational waves}.

\subsection{Vector Perturbations}
\label{subsec:vector-perturb}

From~(\ref{basis-1}), the contraction of~(\ref{basis-3}),
$R^k_{\een\perp k}=0$ and~(\ref{eq:h123}) it follows that
$\varepsilon_\een=0$ and $p_\een=0$.  Raising the index $i$ of
equations~(\ref{basis-2}) with $g_\nul^{ij}$, one finds, using also
that $\Gamma^k_{\nul ij}=\tilde{\Gamma}^k{}_{ij}$,
\begin{equation}\label{basis-2-raise}
  \dot{h}^{kj}_{\perp{|k}}+2Hh^{kj}_{\perp{|k}}=
  2\kappa(\varepsilon_\nul+p_\nul)u^j_{\een},
\end{equation}
where it is used that $\dot{g}^{ij}_\nul=-2Hg^{ij}_\nul$.  It follows
from $R^k_{\een\perp k}=0$,~(\ref{driekrom}) and~(\ref{eq:h123}) that
$h^{ij}_\perp$ must obey
\begin{equation}
  h^{kl}_{\perp |k|l}=0,  \label{eq:hklkl}
\end{equation}
in addition to $h^k_{\perp k}=0$.  Taking the covariant derivative
of~(\ref{basis-2-raise}) with respect to the index $j$ and
using~(\ref{eq:hklkl}) one finds that equations~(\ref{basis-2-raise})
reduce to $u^j_{\een|j}=0$, or, equivalently,
$\boldsymbol{\tilde{\nabla}}\cdot\boldsymbol{u}_\een=0$.  This implies
with~(\ref{fes5}) that $\theta_\een=0$, so that $n_\een=0$.
With~(\ref{eq:nabla-u}) it follows that $\boldsymbol{u}_{\een\perp}$ is
coupled to vector perturbations.  For vector perturbations,
equations~(\ref{subeq:basis}) reduce to the set of equations
\begin{subequations}
  \label{subeq:vector-perturbations}
  \begin{align}
   & \dot{h}^k_{\perp i|k}+2\kappa(\varepsilon_\nul+p_\nul)u_{\een\perp i}=0, \\
   & \ddot{h}^i_{\perp j}+3H\dot{h}^i_{\perp j}+2R^i_{\een\perp j}=0, \\
   & \frac{1}{c}\frac{{\text{d}}}{{\text{d}} t}
    \Bigl[(\varepsilon_\nul+p_\nul) u^i_{\een\perp}\Bigr]+
    5H(\varepsilon_\nul+p_\nul) u^i_{\een\perp}=0.
  \end{align}
\end{subequations}
In view of~(\ref{eq:nabla-u}), vector perturbations are also called
\emph{rotational perturbations}.

\subsection{Scalar Perturbations}

Since both $h^k_{\parallel k}\neq0$ and $R_{\een\parallel}\neq0$
scalar perturbations are coupled to $\varepsilon_\een$, $n_\een$ and
$p_\een$.  It is now demonstrated that $\boldsymbol{u}_{\een\parallel}$ is
coupled to scalar perturbations, by showing that
equations~(\ref{basis-2}) require that the rotation of $\boldsymbol{u}_\een$
vanishes, if the metric tensor is of the form~(\ref{decomp-hij-par}).
Differentiating~(\ref{basis-2}) covariantly with respect to the
index~$j$ and subsequently substituting
expression~(\ref{decomp-hij-par}) yields
\begin{equation}
    2\dot{\phi}_{|i|j}+\dot{\zeta}^{|k}{}_{|k|i|j}-\dot{\zeta}^{|k}{}_{|i|k|j} =
         \kappa c^2(\varepsilon_\nul + p_\nul) u_{\een i|j}.
    \label{eq:feiko1}
\end{equation}
Interchanging~$i$ and~$j$ and subtracting the result
from~(\ref{eq:feiko1}) one gets
\begin{equation}
    \dot{\zeta}^{|k}{}_{|i|k|j}-\dot{\zeta}^{|k}{}_{|j|k|i} =
         -\kappa c^2(\varepsilon_\nul + p_\nul)
         (u_{\een i|j}-u_{\een j|i}),
    \label{dR0i2-rot}
\end{equation}
where it is used that $\dot{\phi}_{|i|j}=\dot{\phi}_{|j|i}$ and
$\dot{\zeta}^{|k}{}_{|k|i|j}=\dot{\zeta}^{|k}{}_{|k|j|i}$. By
rearranging the covariant derivatives, equation~(\ref{dR0i2-rot}) can
be cast in the form
\begin{align}
    (\dot{\zeta}^{|k}{}_{|i|k|j} & -\dot{\zeta}^{|k}{}_{|i|j|k})
    -(\dot{\zeta}^{|k}{}_{|j|k|i}-\dot{\zeta}^{|k}{}_{|j|i|k}) 
   + (\dot{\zeta}^{|k}{}_{|i|j}-\dot{\zeta}^{|k}{}_{|j|i})_{|k} 
        =-\kappa c^2(\varepsilon_\nul + p_\nul)
         (u_{\een i|j}-u_{\een j|i}).
     \label{eq:verwissel}
\end{align}
Using the expressions for the commutator of second-order covariant
derivatives (see \cite{c8}, Chapter~6, Section~5)
\begin{equation}
\label{eq:commu-Riemann}
  A^i{}_{j|p|q}-A^i{}_{j|q|p} =
      A^i{}_k R^k_{\nul jpq} - A^k{}_j R^i_{\nul kpq}, \quad 
   B^i{}_{|p|q}-B^i{}_{|q|p} = B^k R^i_{\nul kpq},
\end{equation}
and substituting the background Riemann tensor~(\ref{eq:glob-curve})
for the three-spaces of constant time one finds that the left-hand
sides of equations~(\ref{eq:verwissel}) vanish identically, implying
that $\boldsymbol{\tilde{\nabla}}\times\boldsymbol{u}_\een=\boldsymbol{0}$.  Therefore, only
$\boldsymbol{u}_{\een\parallel}$ is coupled to scalar perturbations.

\section{Linearized Equations for Scalar Perturbations in Closed,
  Flat and \protect\\ Open FLRW Universes}
\label{sec:scalar-pert}

The true density perturbations $\varepsilon^\true_\een$ and
$n^\true_\een$ are part of the scalar perturbations.  Therefore, a new
set of equations will be derived which describe exclusively scalar
perturbations.  To that end, equations~(\ref{subeq:basis}) will be
rewritten using the results of Section~\ref{sec:decomp-h-u}.

Since scalar perturbations are only coupled to $h^i_{\parallel j}$ and
$u^i_{\een\parallel}$, one may replace
in~(\ref{subeq:basis})--(\ref{fes5}) $h^i{}_j$ by $h^i_{\parallel j}$
and $u^i_\een$ by $u^i_{\een\parallel}$, to obtain perturbation
equations which exclusively describe the evolution of scalar
perturbations.  From now on, only scalar perturbations are considered,
and the subscript $\parallel$ is omitted.  Using the
decompositions~(\ref{eq:h123}), (\ref{eq:prop-Rij}),
(\ref{eq:decomp-u}) and~(\ref{eq:nabla-u}), it will be shown that the
perturbed Einstein equations and conservation laws~(\ref{subeq:basis})
can for scalar perturbations be written in the form
\begin{subequations}
\label{subeq:pertub-flrw}
\begin{align}
  2H(\theta_\een-\vartheta_\een) &=
       \tfrac{1}{2}R_\een + \kappa\varepsilon_\een,
\label{con-sp-1} \\
   \dot{R}_\een &= -2HR_\een+2\kappa
       \varepsilon_\nul(1+w)\vartheta_\een
         -\tfrac{2}{3}R_\nul (\theta_\een-\vartheta_\een), \label{FRW6} \\
   \dot{\varepsilon}_\een &= - 3H(\varepsilon_\een + p_\een)-
         \varepsilon_\nul(1 + w)\theta_\een,  \label{FRW4} \\
   \dot{\vartheta}_\een &= -H(2-3\beta^2)\vartheta_\een-
   \frac{1}{\varepsilon_\nul(1+w)}\dfrac{\tilde{\nabla}^2p_\een}{a^2},
     \quad \beta^2\coloneqq\dfrac{\dot{p}_\nul}{\dot{\varepsilon}_\nul},  \label{FRW5}\\
   \dot{n}_\een &= - 3H n_\een - n_\nul\theta_\een. \label{FRW4a}
\end{align}
\end{subequations}
The system~(\ref{subeq:pertub-flrw}) consists of the energy density
constraint equation or linearized Friedmann equation~(\ref{con-sp-1}),
the momentum constraint equation~(\ref{FRW6}), the energy density
conservation law~(\ref{FRW4}), the momentum conservation
law~(\ref{FRW5}) and the particle number density conservation
law~(\ref{FRW4a}).  As will be shown below, for scalar perturbations
the dynamical equations~(\ref{basis-3}) are automatically fulfilled by
the solution of the system~(\ref{subeq:pertub-flrw}).  This is an
intrinsic property of the Einstein equations and conservation laws.

Note the remarkable similarity between the
systems~(\ref{subeq:einstein-flrw}) and~(\ref{subeq:pertub-flrw}).  In
fact, the system~(\ref{subeq:pertub-flrw}) is the perturbed
counterpart of the system~(\ref{subeq:einstein-flrw}). Just as in the
unperturbed case~(\ref{subeq:einstein-flrw}), the evolution
equations~(\ref{subeq:pertub-flrw}) consist of one algebraic
equation~(\ref{con-sp-1}) and four first-order ordinary differential
equations (\ref{FRW6})--(\ref{FRW4a}) for the five unknown quantities
$\theta_\een$, $R_\een$, $\varepsilon_\een$, $\vartheta_\een$ and
$n_\een$.

Using that
$\dot{p}_\nul=p_n\dot{n}_\nul+p_\varepsilon\dot{\varepsilon}_\nul$ and
the conservation laws~(\ref{FRW2}) and~(\ref{FRW2a}) the quantity
$\beta(t)$ becomes
\begin{equation}
  \label{eq:beta-matter}
  \beta^2=p_\varepsilon+\dfrac{n_\nul p_n}{\varepsilon_\nul(1+w)},
\end{equation}
where $p_\varepsilon$ and $p_n$ are given by~(\ref{eq:p1}).  The
symbol $\tilde{\nabla}^2$ denotes the generalised Laplace operator
with respect to the three-space metric tensor $\tilde{g}_{ij}$,
defined by $\tilde{\nabla}^2 f \coloneqq \tilde{g}^{ij} f_{|i|j}$.

The derivation of the evolution equations~(\ref{subeq:pertub-flrw})
for scalar perturbations is now given.  Firstly, eliminating $\dot{h}^k{}_k$
from~(\ref{basis-1}) with the help of~(\ref{fes5}) yields the
algebraic equation~(\ref{con-sp-1}).

Secondly, multiplying both sides of equations~(\ref{basis-2})
by $g^{ij}_\nul$ and taking the covariant derivative with respect to
the index $j$, one finds
\begin{equation}
  g_\nul^{ij} (\dot{h}^k{}_{k|i|j}-\dot{h}^k{}_{i|k|j}) =
    2\kappa(\varepsilon_\nul+p_\nul) \vartheta_\een,  \label{dR0i4}
\end{equation}
where also~(\ref{fes5}) has been used.  The left-hand side
of~(\ref{dR0i4}) will turn up as a part of the time-derivative of the
curvature $R_\een$.  In fact, differentiating~(\ref{driekrom}) with
respect to time, one gets, using also
$\dot{g}^{ij}_\nul=-2Hg^{ij}_\nul$ and~(\ref{FRW3a}),
\begin{equation}
  \dot{R}_\een = -2HR_\een +
   g_\nul^{ij} (\dot{h}^k{}_{k|i|j}-\dot{h}^k{}_{i|k|j})+
     \tfrac{1}{3}R_\nul \dot{h}^k{}_k,
   \label{dR0i6}
\end{equation}
where it is used that the operations of taking the time-derivative and
the covariant derivative commute, since the background connection
coefficients $\Gamma^k_{\nul ij}=\tilde{\Gamma}^k{}_{ij}$ are
independent of time for \textsc{flrw} metrics~(\ref{eq:metric-flrw}).
Combining~(\ref{dR0i4}) and~(\ref{dR0i6}) and using~(\ref{fes5}) to
eliminate $\dot{h}^k{}_k$ yields~(\ref{FRW6}).  Thus, the
$G^0_{\een i}$ momentum constraint equations~(\ref{basis-2}) have been
recast in one first-order ordinary differential equation~(\ref{FRW6})
for the perturbed Ricci three-scalar, i.e., the local spatial
curvature due to a density perturbation.

Thirdly, it is shown that the dynamical equations~(\ref{basis-3}) are
not needed.  Clearly, the off-diagonal ($i\neq j$)
equations~(\ref{basis-3}) are not coupled to scalar perturbations, so
that these equations need not be taken into account.  For the diagonal
($i=j$) equations, it is sufficient to consider the contraction
of~(\ref{basis-3}).  Eliminating the quantity $\dot{h}^k{}_k$ with the
help of~(\ref{fes5}), one arrives at
\begin{equation}
   \dot{\theta}_\een-\dot{\vartheta}_\een +
      6H(\theta_\een-\vartheta_\een)-R_\een=
       \tfrac{3}{2}\kappa(\varepsilon_\een-p_\een).
        \label{eq:168a}
\end{equation}
Using~(\ref{con-sp-1}) to eliminate the second term of~(\ref{eq:168a})
yields for the contraction of~(\ref{basis-3})
\begin{equation}
  \label{eq:168a-kort}
  \dot{\theta}_\een-\dot{\vartheta}_\een + \tfrac{1}{2}R_\een=
       -\tfrac{3}{2}\kappa(\varepsilon_\een+p_\een).
\end{equation}
This equation is identical to the time-derivative of the constraint
equation~(\ref{con-sp-1}).  Differentiation of
equation~(\ref{con-sp-1}) with respect to time yields
\begin{equation}
\label{eq:con-sp-1-sum-time}
2\dot{H}(\theta_\een-\vartheta_\een)+2H(\dot{\theta}_\een-\dot{\vartheta}_\een)
 =\tfrac{1}{2}\dot{R}_\een+\kappa\dot{\varepsilon}_\een.
\end{equation}
Eliminating the time-derivatives $\dot{H}$, $\dot{R}_\een$ and
$\dot{\varepsilon}_\een$ with the help of (\ref{FRW3})--(\ref{FRW2}),
(\ref{FRW6}) and~(\ref{FRW4}), respectively, yields the dynamical
equation~(\ref{eq:168a-kort}).  Consequently, for scalar perturbations
the dynamical equations~(\ref{basis-3}) are not needed.

Finally, taking the covariant derivative of~(\ref{basis-5}) with
respect to the background metric tensor $g_{\nul ij}$ and
using~(\ref{fes5}), one gets
\begin{equation}
  \frac{1}{c}\frac{\text{d}}{\text{d}t}
     \Bigl[(\varepsilon_\nul+p_\nul)\vartheta_\een\Bigr]-
    g^{ik}_\nul p_{\een|k|i}+5H(\varepsilon_\nul+p_\nul)\vartheta_\een = 0,  
\label{mom2}
\end{equation}
where it is used that the operations of taking the time-derivative and
the covariant derivative commute.  With~(\ref{eq:metric-flrw}),
$\tilde{\nabla}^2f\coloneqq\tilde{g}^{ij}f_{|i|j}$ and~(\ref{FRW2})
one can rewrite~(\ref{mom2}) in the form
\begin{equation}
  \dot{\vartheta}_\een +
 H\left(2-3\frac{\dot{p}_\nul}{\dot{\varepsilon}_\nul}\right)\vartheta_\een+
     \frac{1}{\varepsilon_\nul+p_\nul}\dfrac{\tilde{\nabla}^2 p_\een}{a^2}=0.
                \label{mom3}
\end{equation}
Using the definitions $w\coloneqq p_\nul/\varepsilon_\nul$ and
$\beta^2\coloneqq\dot{p}_\nul/\dot{\varepsilon}_\nul$ one arrives at
equation~(\ref{FRW5}).

The decomposition theorems of York, Stewart, and Walker have been used
to rewrite the full system of Einstein equations and conservation
laws~(\ref{subeq:basis}) into a substantially simpler system of
equations~(\ref{subeq:pertub-flrw}) which describes exclusively scalar
perturbations.

With the derivation of equations~(\ref{subeq:pertub-flrw}) the first
step towards the solution of the gauge problem of cosmology has been
taken.  In the next section, the background
equations~(\ref{subeq:einstein-flrw}) and the linearized
equations~(\ref{subeq:pertub-flrw}) will be used to identify which
gauge-invariant quantities could be equal to $\varepsilon^\true_\een$
and $n^\true_\een$.

\section{Unique Gauge-invariant Quantities}
\label{sec:unique}

The background equations~(\ref{subeq:einstein-flrw}) and the
perturbation equations~(\ref{subeq:pertub-flrw}) are both written with
respect to the \emph{same} system of reference. Therefore, these two
systems can be combined to describe the evolution of scalar
perturbations.  The system~(\ref{subeq:einstein-flrw}) describes the
evolution of the quantities $\theta_\nul=3H$, $R_\nul$,
$\varepsilon_\nul$, $\vartheta_\nul$ and~$n_\nul$, whereas the
system~(\ref{subeq:pertub-flrw}) describes the evolution of their
perturbed counterparts $\theta_\een$, $R_\een$, $\varepsilon_\een$,
$\vartheta_\een$ and~$n_\een$.  The quantities $\varepsilon$, $n$
and~$\theta$ (\ref{eq:scalars-flrw}) are scalars. Let~$S$ represent
one of the three scalars $\varepsilon$, $n$ and $\theta$, then a
perturbation to these scalars transforms under a \emph{general}
infinitesimal coordinate transformation~(\ref{func}) as
\begin{equation}
  \label{eq:trans-scalar}
  S_\een(t,\boldsymbol{x}) \rightarrow S^\prime_\een= S_\een(t,\boldsymbol{x})+\xi^0(t,\boldsymbol{x})\dot{S}_\nul(t),
\end{equation}
where $S_\nul$ and $S_\een$ are the background and the perturbation,
respectively, of one of these scalars. The term
$\hat{S}_\een\coloneqq \mathcal{L}_\xi S_\nul=\xi^0\dot{S}_\nul$ is
the gauge mode, where $\mathcal{L}_\xi$ is the Lie-derivative operator
with respect to the infinitesimal four-vector~$\xi^\mu$. It can be
easily verified by substitution that the complete set of gauge modes
for the system of equations~(\ref{subeq:pertub-flrw}) is
\begin{subequations}
\label{subeq:gauge-dep}
\begin{align}
 & \hat{\varepsilon}_\een = 
        \psi\dot{\varepsilon}_\nul, \quad
     \hat{n}_\een  = 
        \psi\dot{n}_\nul, \quad
     \hat{\theta}_\een  =
        \psi\dot{\theta}_\nul, \label{tr-theta}\\
   &     \hat{\vartheta}_\een  =-\frac{\tilde{\nabla}^2\psi}{a^2}, \quad
       \hat{R}_\een  =  
     4H\left[\frac{\tilde{\nabla}^2\psi}{a^2} -
       \tfrac{1}{2}R_\nul\psi\right],
            \label{drie-ijk}
\end{align}
\end{subequations}
where $\xi^0=\psi(\boldsymbol{x})$ in synchronous coordinates,
see~(\ref{eq:synchronous}).  The quantities~(\ref{subeq:gauge-dep})
are mere coordinate artefacts, which have no physical meaning, since
the gauge function $\psi(\boldsymbol{x})$ is an arbitrary
infinitesimal function.  Since the set~(\ref{subeq:gauge-dep}) is a
solution set of the system of \emph{linear}
equations~(\ref{subeq:pertub-flrw}), one can generate new, equivalent,
solutions by adding the gauge modes~(\ref{subeq:gauge-dep}) to the
original solutions $\varepsilon_\een$, $n_\een$, $\theta_\een$,
$\vartheta_\een$ and $R_\een$.  Therefore, the solutions of the
system~(\ref{subeq:pertub-flrw}) have no physical significance. This
is the well-known gauge problem of cosmology.  A first step towards
the solution of the gauge problem has already been taken in
Section~\ref{sec:scalar-pert}, by rewriting the system of
equations~(\ref{subeq:basis}) into a substantially smaller
system~(\ref{subeq:pertub-flrw}) that describes exclusively scalar
perturbations.  This has reduced the number of gauge-dependent
quantities to five.  The second step towards the solution of the gauge
problem of cosmology can now be taken.  Since $\varepsilon_\een$,
$n_\een$ and~$\theta_\een$ transform under the general infinitesimal
transformation~(\ref{func}) according to~(\ref{eq:trans-scalar}), one
can combine two independent scalars to eliminate the gauge function
$\xi^0(t,\boldsymbol{x})$. With the three independent
scalars~(\ref{eq:scalars-flrw}), one can construct ${3\choose2}=3$
different sets of three gauge-invariant quantities.  In each of these
sets exactly one gauge-invariant quantity vanishes.  Therefore, the
only non-trivial choice is
\begin{subequations}
  \label{subeq:gi-quant}
  \begin{align}
   & \varepsilon_\een^{\text{gi}}
        \coloneqq\varepsilon_\een-\dfrac{\dot{\varepsilon}_\nul}
        {\dot{\theta}_\nul}\theta_\een, \quad
   n_\een^{\text{gi}}\coloneqq n_\een-\dfrac{\dot{n}_\nul}
        {\dot{\theta}_\nul}\theta_\een, \label{e-n-gi} \\
      &   \theta_\een^{\text{gi}} \coloneqq\theta_\een-\dfrac{\dot{\theta}_\nul}
        {\dot{\theta}_\nul}\theta_\een=0. \label{theta-gi}
  \end{align}
\end{subequations}
It follows from the transformation rule~(\ref{eq:trans-scalar}) that
the quantities~(\ref{subeq:gi-quant}) are indeed invariant under the
general infinitesimal transformation~(\ref{func}), i.e., they are
\emph{gauge-invariant}, hence the superscript~`gi'. Since the
evolution of \textsc{flrw} universes is determined by only three
independent scalars~(\ref{eq:scalars-flrw}), the
quantities~(\ref{e-n-gi}) are the only non-trivial gauge-invariant
quantities, i.e., these quantities are \emph{unique}.  From this fact
one may draw the conclusion that the quantities~(\ref{e-n-gi}) are the
true, physical density perturbations, i.e.,
$\varepsilon_\een^{\text{gi}}\equiv\varepsilon_\een^\true$ and
$n_\een^{\text{gi}}\equiv n_\een^\true$, and that the
quantity~(\ref{theta-gi})
$\theta_\een^{\text{gi}}\equiv\theta_\een^\true=0$ is the true
physical perturbation to the global expansion $\theta_\nul$ of the
universe. However, to be sure that this is indeed the case, one needs
to study the system~(\ref{subeq:pertub-flrw}) together with the
quantities~(\ref{subeq:gi-quant}) in the non-relativistic limit.  This
third and last step will be taken in the next section.

\section{Non-relativistic Limit}
\label{sec:newt-limit}

The standard non-relativistic limit is defined by four requirements,
see, e.g., \cite{carroll-2003}, page~153:
\begin{enumerate}
\item The gravitational field should be weak, i.e., can be considered
  as a perturbation of a flat \textsc{flrw} universe.
\item The particles are moving slowly with respect to the speed of
  light.
\item The transformation~(\ref{func}) with~(\ref{eq:synchronous}) is
  limited to transformations for which the Newtonian Theory of Gravity
  in invariant.
\item The gravitational field of a density perturbation should be
  static, i.e., it does not change with time.
\end{enumerate}
Requirement~3, which is not used in the literature, is needed because
one has to deal with gauge transformations~(\ref{func})
with~(\ref{eq:synchronous}) in an expanding universe.

For cosmological perturbations the gravitational field is already
weak.  In order to meet the first requirement a flat, $R_\nul=0$,
\textsc{flrw} universe is considered.  In a flat \textsc{flrw}
universe the expressions~(\ref{e-n-gi}) reduce to
\begin{subequations}
  \label{eq:e-n-gi-flat}
  \begin{align}
    \varepsilon_\een^{\text{gi}} & = -\frac{1}{\kappa}(2H\vartheta_\een+\tfrac{1}{2}R_\een),
             \label{eq:e-n-gi-flat-a}\\
  n_\een^{\text{gi}} & =n_\een-\frac{n_\nul(\kappa\varepsilon_\een+2H\vartheta_\een+
    \tfrac{1}{2}R_\een)}{\kappa\varepsilon_\nul(1+w)}, \label{eq:e-n-gi-flat-b}
  \end{align}
\end{subequations}
where the background equations~(\ref{subeq:einstein-flrw}) have been
used to eliminate all time-derivatives, the
constraint equation~(\ref{con-sp-1}) to eliminate~$\theta_\een$, and $3H=\theta_\nul$,

Using~(\ref{decomp-hij-par}) and the fact that spatial covariant
derivatives become in flat three-space ordinary derivatives with
respect to the spatial coordinates, the local perturbation to the
spatial curvature,~(\ref{driekrom}) and its gauge
mode~(\ref{drie-ijk}), reduce for a flat \textsc{flrw} universe~to
\begin{equation}\label{RnabEE-0}
   R_\een =\dfrac{4}{c^2}\phi^{|k}{}_{|k}=
       -\dfrac{4}{c^2}\dfrac{\nabla^2\phi}{a^2}, \quad \hat{R}_\een=4H\dfrac{\nabla^2\psi}{a^2},
\end{equation}
where $\nabla^2$ is the usual Laplace operator.  Substituting the
expression for $R_\een$ into the perturbation
equations~(\ref{subeq:pertub-flrw}), one gets
\begin{subequations}
\label{subeq:pertub-gi-flat}
\begin{align}
     H(\theta_\een-\vartheta_\een)
        &=-\dfrac{1}{c^2}\dfrac{\nabla^2\phi}{a^2}+ \dfrac{4\pi G_{\text{N}}}{c^4}
        \left[\varepsilon^{\text{gi}}_\een+
         \dfrac{\dot{\varepsilon}_\nul}{\dot{\theta}_\nul}\theta_\een\right],
\label{con-sp-1-flat} \\
     \dfrac{\nabla^2\dot{\phi}}{a^2} &= -\dfrac{4\pi G_{\text{N}}}{c^2}
\varepsilon_\nul(1 + w)\vartheta_\een, \label{FRW6gi-flat} \\
    \dot{\varepsilon}_\een &=-3H(\varepsilon_\een + p_\een)-
         \varepsilon_\nul(1 + w)\theta_\een,  \label{FRW4gi-flat} \\
  \dot{\vartheta}_\een &= -H(2-3\beta^2)\vartheta_\een-
   \frac{1}{\varepsilon_\nul(1+w)}\dfrac{\nabla^2p_\een}{a^2}, 
  \label{FRW5gi-flat} \\
  \dot{n}_\een &= - 3H n_\een -
         n_\nul\theta_\een, \label{FRW4agi-flat}
\end{align}
\end{subequations}
where~(\ref{e-n-gi}) has been used to eliminate $\varepsilon_\een$
from the $G^0_{\een 0}$ constraint equation~(\ref{con-sp-1}), and
$H\coloneqq\dot{a}/a$.  With $R_\nul=0$ the first requirement is satisfied.

Next, the second requirement is implemented.  Since the spatial parts
$u^i_\een$ of the fluid four-velocity are gauge-dependent~(\ref{drie-ijk}),
the second requirement must be defined by\footnote{Recall that
  $\boldsymbol{u}_\een=\boldsymbol{u}_{\een\parallel}$ is the \emph{irrotational} part
  of the three-space fluid velocity. The rotational part
  $\boldsymbol{u}_{\een\perp}$ is not coupled to density perturbations and need,
  therefore, not be considered. See Section~\ref{sec:decomp-h-u}.}
\begin{equation}
  \label{eq:non-rel-lim}
  u^i_{\een\text{phys}} \coloneqq
    \dfrac{1}{c}U^i_{\een\true} \rightarrow 0,
\end{equation}
i.e., the \emph{physical} spatial parts of the fluid
four-velocity are negligible with respect to the speed of
light. With~(\ref{eq:non-rel-lim}) the second requirement is satisfied.

Next, the third requirement will be addressed.  In the
limit~(\ref{eq:non-rel-lim}), the mean kinetic energy per particle
$\tfrac{1}{2}m\langle{v^2}\rangle\rightarrow0$ is very small compared
to the rest energy $mc^2$ per particle.  This implies that the
pressure $p\rightarrow0$ is vanishingly small with respect to the rest
energy density $nmc^2$.  Substituting $p=0$, i.e., $p_\nul=0$ and
$p_\een=0$, into the momentum conservation laws~(\ref{basis-5})
yields, using also the background equation~(\ref{FRW2}) with
$w\coloneqq p_\nul/\varepsilon_\nul\rightarrow0$,
\begin{equation}\label{eq:ui-par-p0}
    \dot{u}^i_\een=-2H u^i_\een.
\end{equation}
Since the components $u^i_{\een\text{phys}}$ vanish in the
non-relativistic limit~(\ref{eq:non-rel-lim}), the general solutions
of equations~(\ref{eq:ui-par-p0}) are equal to the gauge modes,
see~(\ref{drie-ijk}),
\begin{equation}\label{eq:ui-gauge-mode}
    \hat{u}^{i}_\een(t,\boldsymbol{x})=-\dfrac{1}{a^2(t)}
\tilde{g}^{ik}(\boldsymbol{x})\dfrac{\partial\psi(\boldsymbol{x})}{\partial x^k},
\end{equation}
where it is used that $H\coloneqq\dot{a}/a$.  As a result, in the
limit~(\ref{eq:non-rel-lim}) one is left with the gauge
modes~(\ref{eq:ui-gauge-mode}) only.  Since $\hat{u}^i_\een=0$ are
solutions of~(\ref{eq:ui-par-p0}), one may, without losing any
physical information, put the gauge modes $\hat{u}^i_\een$ equal to
zero, i.e.,
\begin{equation}
  \label{eq:gauge-ui-nul}
     \hat{u}^i_\een=0.
\end{equation}
From~(\ref{eq:ui-gauge-mode}) and~(\ref{eq:gauge-ui-nul}) it now
follows that $\partial\psi/\partial x^k=0$, so that $\psi=C$ is an
arbitrary constant in the non-relativistic limit.  Substituting
$\psi=C$ into~(\ref{eq:synchronous}) one finds that the relativistic
transformation~(\ref{func}) between synchronous coordinates reduces in
the limit~(\ref{eq:non-rel-lim}) to the infinitesimal transformation
\begin{equation}
\label{eq:gauge-trans-newt}
x^0 \rightarrow x^{\prime0}=x^0 - C, \quad
        x^i \rightarrow x^{\prime i}= x^i-\chi^i(\boldsymbol{x}),
\end{equation}
where $\chi^i(\boldsymbol{x})$ are three arbitrary functions of the spatial
coordinates.  Thus, in the non-relativistic limit time and space
transformations are decoupled: time coordinates may be shifted and
spatial coordinates may be chosen arbitrarily.  The residual gauge
freedoms $C$ and $\chi^i(\boldsymbol{x})$ in the non-relativistic limit
express the fact that the Newtonian Theory of Gravity is invariant
under the gauge transformation~(\ref{eq:gauge-trans-newt}). This
satisfies the third requirement.

Next, the fourth and last requirement will be discussed.
Using~(\ref{eq:non-rel-lim}) and~(\ref{eq:gauge-ui-nul}), on gets
\begin{equation}
  \label{eq:div-u-nul}
  \vartheta_\een:=u^k_{\een|k}=0.
\end{equation}
Using~(\ref{eq:div-u-nul}) and $p=0$, the system of Einstein equations
and conservation laws~(\ref{subeq:pertub-gi-flat}) becomes in the
non-relativistic~limit:
\begin{subequations}
\label{subeq:pertub-gi-flat-newt}
\begin{align}
   \nabla^2\phi &= \dfrac{4\pi G_{\text{N}}}{c^2}a^2\varepsilon^{\text{gi}}_\een,
         \label{con-sp-1-flat-newt} \\
   \nabla^2\dot{\phi} &= 0, \label{FRW6gi-flat-newt} \\
  \dot{\varepsilon}_\een &= - 3H\varepsilon_\een-\varepsilon_\nul\theta_\een,
                           \label{FRW4gi-flat-newt} \\
   \dot{n}_\een &= - 3H n_\een-n_\nul\theta_\een, \label{FRW4agi-flat-newt}
\end{align}
\end{subequations}
The $G^0_{\een0}$ constraint equation~(\ref{con-sp-1-flat-newt}) can
be found by observing that the time-derivative of the background
constraint equation, or Friedmann equation~(\ref{FRW3}), multiplied by
$\tfrac{1}{6}\theta_\een/\dot{H}$, reads
\begin{equation}
  \label{eq:time-deriv-constraint}
  H\theta_\een=\dfrac{4\pi G_{\text{N}}}{c^4}
       \dfrac{\dot{\varepsilon}_\nul}{\dot{\theta}_\nul}\theta_\een,
\end{equation}
where it is used that $R_\nul=0$ and $\theta_\nul=3H$.
Combining~(\ref{con-sp-1-flat}), (\ref{eq:div-u-nul})
and~(\ref{eq:time-deriv-constraint}) yields
equation~(\ref{con-sp-1-flat-newt}).

The right-hand side of~(\ref{con-sp-1-flat-newt}) is clearly
gauge-invariant. Since the gauge mode
$\hat{R}_\een=0$,~(\ref{RnabEE-0}), $R_\een$ is gauge-invariant, so
that the left-hand side of~(\ref{con-sp-1-flat-newt}) is also
gauge-invariant.

Due to~(\ref{eq:div-u-nul}) there is no fluid flow so that density
perturbations do not evolve.  This, combined with~(\ref{theta-gi}),
implies the basic fact of the Newtonian Theory of Gravity, namely that
the gravitational field is \emph{static}, as follows from the momentum
constraint equation~(\ref{FRW6gi-flat-newt}).  Consequently,
$a^2(t)\varepsilon^{\text{gi}}_\een(t,\boldsymbol{x})$
in~(\ref{con-sp-1-flat-newt}) should be replaced by
$a^2(t_0)\varepsilon^{\text{gi}}_\een(\boldsymbol{x})$.  Defining the
potential
$\varphi(\boldsymbol{x})\coloneqq\phi(\boldsymbol{x})/a^2(t_0)$,
equations~(\ref{con-sp-1-flat-newt}) and~(\ref{FRW6gi-flat-newt})
imply
\begin{equation}
  \label{eq:poisson}
  \nabla^2\varphi(\boldsymbol{x})=4\pi G_{\text{N}} \rho_\een(\boldsymbol{x}), \quad
    \rho_\een(\boldsymbol{x})\coloneqq\dfrac{\varepsilon^{\text{gi}}_\een(\boldsymbol{x})}{c^2},
\end{equation}
which is the Poisson equation of the Newtonian Theory of Gravity.
With~(\ref{eq:poisson}) the fourth requirement for the
non-relativistic limit, i.e., a static gravitational field, has been
satisfied.

Finally, expressions~(\ref{eq:e-n-gi-flat}) will be checked.
Using~(\ref{eq:div-u-nul}), expression~(\ref{eq:e-n-gi-flat-a})
reduces to $\varepsilon^{\text{gi}}_\een=-R_\een/2\kappa$, which is,
with~(\ref{RnabEE-0}) and~(\ref{FRW6gi-flat-newt}), equal to the
Poisson equation~(\ref{eq:poisson}).  The background
equations~(\ref{subeq:einstein-flrw}) are in the non-relativistic
limit
\begin{equation}
  \label{eq:einstein-flrw-nonrel}
  3H^2=\kappa\varepsilon_\nul+\Lambda, \quad
  \dot{\varepsilon}_\nul=-3H\varepsilon_\nul, \quad \dot{n}_\nul=-3H n_\nul.
\end{equation}
From equations~(\ref{FRW4gi-flat-newt})--(\ref{FRW4agi-flat-newt})
and~(\ref{eq:einstein-flrw-nonrel}) it follows that
$\varepsilon_\een=n_\een mc^2$ and $\varepsilon_\nul=n_\nul mc^2$,
respectively, so that expression~(\ref{eq:e-n-gi-flat-b}) becomes,
using that $R_\een=-2\kappa\varepsilon^{\text{gi}}_\een$, in a
pressure-less fluid
\begin{equation}
  \label{eq:newt-ngi}
    n^{\text{gi}}_\een(\boldsymbol{x})=\dfrac{\varepsilon^{\text{gi}}_\een(\boldsymbol{x})}{mc^2},
   \quad \rho_\een(\boldsymbol{x})=mn^{\text{gi}}_\een(\boldsymbol{x}),
\end{equation}
where the first expression is a well-known result of the Special
Theory of Relativity.

Before a conclusion can be drawn, a few more remarks are made.

From~(\ref{eq:div-u-nul}) and~(\ref{FRW6gi-flat-newt}) it follows that
the perturbed expansion scalar~(\ref{fes5}) and its
gauge mode~(\ref{tr-theta}) are in the non-relativistic limit given by
\begin{equation}
  \label{eq:theta-1-nrl}
  \theta_\een=-\dfrac{1}{c^2}\dot{\zeta}^{|k}{}_{|k}, \quad \hat{\theta}_\een=C\dot{\theta}_\nul,
\end{equation}
where~(\ref{decomp-hij-par}) has been used.  Using that
$\theta_\nul=3H$, it follows from~(\ref{eq:einstein-flrw-nonrel})
and~(\ref{eq:theta-1-nrl}) that $\dot{\theta}_\nul\neq0$ and
$\theta_\een\neq0$, so that the quantities~(\ref{subeq:gi-quant}) are
well-defined.  Consequently, it follows from~(\ref{e-n-gi}) that
\begin{equation}
  \label{eq:notequal-in-nrl}
\varepsilon_\een^{\text{gi}}\neq\varepsilon_\een, \quad n_\een^{\text{gi}}\neq n_\een,
\end{equation}
independent of the scale of a perturbation.

Equations~(\ref{FRW4gi-flat-newt}) and~(\ref{FRW4agi-flat-newt}) are
leftovers of the conservation laws~(\ref{FRW4gi-flat})
and~(\ref{FRW4agi-flat}) in the non-relativistic limit.  They have
only the gauge modes~(\ref{tr-theta})
$\hat{\varepsilon}_\een=C\dot{\varepsilon}_\nul$,
$\hat{n}_\een=C\dot{n}_\nul$ and
$\hat{\theta}_\een=C\dot{\theta}_\nul$ as solutions and have, as a
consequence, no physical significance.  These gauge modes never become
zero, since the universe is not static in the non-relativistic limit,
as follows from the Friedmann equation and conservation
laws~(\ref{eq:einstein-flrw-nonrel}).  Since
equations~(\ref{FRW4gi-flat-newt}) and~(\ref{FRW4agi-flat-newt}) are
decoupled from the physical equations~(\ref{con-sp-1-flat-newt})
and~(\ref{FRW6gi-flat-newt}) they are not part of the Newtonian Theory
of Gravity and need, therefore, not be considered.  Consequently, in
the non-relativistic limit one is left with the Poisson
equation~(\ref{eq:poisson}) and the relation~(\ref{eq:newt-ngi}).

This section will be closed with an important conclusion.  It has been
shown that the complete set of linearized Einstein equations and
conservation laws for scalar perturbations~(\ref{subeq:pertub-flrw}),
combined with the gauge-invariant quantities~(\ref{e-n-gi}) reduce in
the non-relativistic limit to the Newtonian
equations~(\ref{eq:poisson}) and~(\ref{eq:newt-ngi}).  This implies
that the gauge-invariant quantities $\varepsilon_\een^{\text{gi}}$ and
$n_\een^{\text{gi}}$ are equal to the true, physical perturbations
$\varepsilon^\true_\een$ and $n^\true_\een$, respectively, in the
non-relativistic limit.  It must, therefore, be concluded that
$\varepsilon_\een^{\text{gi}}$ and $n_\een^{\text{gi}}$ are also equal
to the true physical density perturbations $\varepsilon^\true_\een$
and $n^\true_\een$ in the General Theory of Relativity, i.e.,
\begin{equation}
\varepsilon_\een^{\text{gi}}
        \coloneqq\varepsilon_\een-\dfrac{\dot{\varepsilon}_\nul}
        {\dot{\theta}_\nul}\theta_\een \equiv \varepsilon^\true_\een, \quad
   n_\een^{\text{gi}}\coloneqq n_\een-\dfrac{\dot{n}_\nul}
        {\dot{\theta}_\nul}\theta_\een\equiv n^\true_\een.
  \label{eq:e-true-and-n-true}
\end{equation}
In view of the results in this section,~(\ref{theta-gi}) can be written as
\begin{equation}
  \label{eq:theta-phys}
  \theta_\een^{\text{gi}} \equiv \theta_\een^\true=0,
\end{equation}
implying that, in the linear regime, the \emph{global} expansion
$\theta_\nul=3H$ is not affected by a \emph{local} density
perturbation.

With~(\ref{eq:e-true-and-n-true}) and~(\ref{eq:theta-phys}) the
gauge problem of cosmology for closed, flat and open \textsc{flrw}
universes is solved.

\section{Evolution Equations for the Density
  Contrast Functions}
\label{sec:flat-pert}

The evolution of the density
perturbations~(\ref{eq:e-true-and-n-true}) is completely determined by
the background equations~(\ref{subeq:einstein-flrw}) and their
perturbed counterparts~(\ref{subeq:pertub-flrw}).  In principle, these
two systems can be used to study the evolution of density
perturbations in \textsc{flrw} universes.  However, the
system~(\ref{subeq:pertub-flrw}) is still too complicated, since it
also admits the non-physical solutions~(\ref{subeq:gauge-dep}).
Therefore, evolution equations for $\varepsilon^\true_\een$ and
$n^\true_\een$ are derived.

Firstly, it is observed that the gauge-dependent quantity
$\theta_\een$ is not needed in the calculations, since the local
perturbation to the global expansion,
$\theta^\true_\een$~(\ref{eq:theta-phys}), vanishes identically.
Eliminating $\theta_\een$ from the differential equations
(\ref{FRW6})--(\ref{FRW4a}) with the help of the algebraic
constraint equation~(\ref{con-sp-1}) yields a system of four
first-order ordinary differential equations
\begin{subequations}
\label{subeq:pertub-gi}
\begin{align}
   \dot{\varepsilon}_\een + 3H(\varepsilon_\een +p_\een)
      +\varepsilon_\nul(1 + w)\Bigl[\vartheta_\een +\frac{1}{2H}\left(
 \kappa\varepsilon_\een+\tfrac{1}{2}R_\een\right)\Bigr] &= 0,
       \label{FRW4gi} \\
   \dot{n}_\een + 3H n_\een +
       n_\nul \Bigl[\vartheta_\een +\frac{1}{2H}\left(\kappa\varepsilon_\een +
      \tfrac{1}{2}R_\een\right)\Bigr] &= 0, \label{FRW4agi} \\
  \dot{\vartheta}_\een + H(2-3\beta^2)\vartheta_\een +
        \frac{1}{\varepsilon_\nul(1+w)}
      \dfrac{\tilde{\nabla}^2p_\een}{a^2} &= 0,
\label{FRW5gi}\\
  \dot{R}_\een
     +2HR_\een - 2\kappa\varepsilon_\nul(1 + w)\vartheta_\een
     +\frac{R_\nul}{3H} \left(\kappa\varepsilon_\een +
 \tfrac{1}{2}R_\een \right) & =0,  \label{FRW6gi} 
\end{align}
\end{subequations}
for the four quantities $\varepsilon_\een$, $n_\een$, $\vartheta_\een$
and~$R_\een$.

Using the background equations~(\ref{subeq:einstein-flrw}) to
eliminate all time-derivatives and the linearized constraint
equation~(\ref{con-sp-1}) to eliminate~$\theta_\een$, the density
perturbations~(\ref{eq:e-true-and-n-true}) become
\begin{subequations}
\label{subeq:pertub-gi-e-n}
\begin{align}
   & \varepsilon^\true_\een  =
     \dfrac{ \varepsilon_\een R_\nul -
   3\varepsilon_\nul(1 + w) (2H\vartheta_\een +
  \tfrac{1}{2}R_\een) }
  { R_\nul+3\kappa\varepsilon_\nul(1 + w)},
        \label{Egi} \\
   & n^\true_\een  = n_\een-\dfrac{3n_\nul(\kappa\varepsilon_\een+2H\vartheta_\een+
            \tfrac{1}{2}R_\een)}
         {R_\nul+3\kappa\varepsilon_\nul(1+w)}.  \label{nu2}
\end{align}
\end{subequations}
The evolution of these density perturbations is completely determined
by the background equations~(\ref{subeq:einstein-flrw}) and the set of
first-order differential equations~(\ref{subeq:pertub-gi}).  In the
study of the evolution of density perturbations, it is convenient not
to use $\varepsilon^\true_{\een}$ and $n^\true_{\een}$, but instead
their corresponding contrast functions $\delta_\varepsilon$ and
$\delta_n$, defined by
\begin{equation}\label{eq:contrast}
  \delta_\varepsilon(t,\boldsymbol{x}) \coloneqq
      \dfrac{\varepsilon^\true_\een(t,\boldsymbol{x})}{\varepsilon_\nul(t)}, \quad
  \delta_n(t,\boldsymbol{x}) \coloneqq
      \dfrac{n^\true_\een(t,\boldsymbol{x})}{n_\nul(t)}.
\end{equation}
The system of equations~(\ref{subeq:pertub-gi}) for the four
independent quantities $\varepsilon_\een$, $n_\een$, $\vartheta_\een$
and~$R_\een$ is now recast, using the procedure given in the appendix,
in a new system of equations for the two independent contrast
functions $\delta_\varepsilon$ and $\delta_n$.  In this procedure it
is explicitly assumed that $p\neq0$, i.e., the pressure does not
vanish identically.  The case $p\rightarrow0$ has been considered in
Section~\ref{sec:newt-limit} on the non-relativistic limit.  The final
results are the evolution equations for the relative density
perturbations~(\ref{eq:contrast}) in \textsc{flrw} universes:
\begin{subequations}
\label{subeq:final}
\begin{align}
   \ddot{\delta}_\varepsilon + b_1 \dot{\delta}_\varepsilon +
      b_2 \delta_\varepsilon &=
      b_3 \left[\delta_n - \frac{\delta_\varepsilon}{1+w}\right],
              \label{sec-ord}  \\
   \frac{1}{c}\frac{{\text{d}}}{{\text{d}} t}
      \left[\delta_n - \frac{\delta_\varepsilon}{1 + w}\right] & =
     \frac{3Hn_\nul p_n}{\varepsilon_\nul(1 + w)}
     \left[\delta_n - \frac{\delta_\varepsilon}{1 + w}\right],
                 \label{fir-ord}
\end{align}
\end{subequations}
where the coefficients $b_1$, $b_2$ and $b_3$ are given by
\begin{subequations}
\label{subeq:coeff-contrast}
 \begin{align}
  b_1  = &\, \dfrac{\kappa\varepsilon_\nul(1+w)}{H}
  -2\dfrac{\dot{\beta}}{\beta}-H(2+6w+3\beta^2)
   + R_\nul\left[\dfrac{1}{3H}+
  \dfrac{2H(1+3\beta^2)}
  {R_\nul+3\kappa\varepsilon_\nul(1+w)}\right], \\
  b_2 = & -\tfrac{1}{2}\kappa\varepsilon_\nul(1+w)(1+3w) 
  H^2\left[1-3w+6\beta^2(2+3w)\right]
   +6H\dfrac{\dot{\beta}}{\beta}\left[w+\dfrac{\kappa\varepsilon_\nul(1+w)}
   {R_\nul+3\kappa\varepsilon_\nul(1+w)}\right]  \nonumber \\
   & - R_\nul\left(\tfrac{1}{2}w+
\dfrac{H^2(1+6w)(1+3\beta^2)}{R_\nul+3\kappa\varepsilon_\nul(1+w)}
\right)
-\beta^2\left(\frac{\tilde{\nabla}^2}{a^2}-\tfrac{1}{2}R_\nul
\right), \\
  b_3  =&\,
\Biggl\{\dfrac{-18H^2}{R_\nul+3\kappa\varepsilon_\nul(1+w)}
  \Biggl[\varepsilon_\nul p_{\varepsilon n}(1+w) 
   +\dfrac{2p_n}{3H}\dfrac{\dot{\beta}}{\beta}
    +p_n(p_\varepsilon-\beta^2)+n_\nul p_{nn}\Biggr]+
   p_n\Biggr\}\dfrac{n_\nul}{\varepsilon_\nul}
\left(\frac{\tilde{\nabla}^2}{a^2}-\tfrac{1}{2}R_\nul\right).
\label{eq:b3}
\end{align}
\end{subequations}
In these expressions the partial derivatives of the pressure,
$p_\varepsilon$ and $p_n$, are defined by~(\ref{eq:p1}). The second-order partial derivatives are given by
\begin{equation}
  \label{eq:sec-ord-partial}
  p_{nn}\coloneqq\dfrac{\partial^2p}{\partial n^2}, \quad
   p_{\varepsilon n}\coloneqq\dfrac{\partial^2p}{\partial\varepsilon\,\partial n}.
\end{equation}
In the derivation of the coefficients~(\ref{subeq:coeff-contrast}) it
is used that the time-derivative of $w$ is
\begin{equation}
  \label{eq:time-w}
  \dot{w}=3H(1+w)(w-\beta^2).
\end{equation}
This relation follows from the definitions
$w\coloneqq p_\nul/\varepsilon_\nul$ and
$\beta^2\coloneqq\dot{p}_\nul/\dot{\varepsilon}_\nul$ using only the
energy conservation law~(\ref{FRW2}), and is, therefore,
\emph{independent} of the precise form of the equation of state.

The system of equations~(\ref{subeq:final}) is equivalent to a system
of \emph{three} first-order differential equations, whereas the
original system~(\ref{subeq:pertub-gi}) is a \emph{fourth}-order
system. This difference is due to the fact that the gauge
modes~(\ref{subeq:gauge-dep}), which are solutions of the
system~(\ref{subeq:pertub-gi}), are completely removed from the solution
set of~(\ref{subeq:final}): one degree of freedom, namely the unknown
gauge function $\xi^0(t,\boldsymbol{x})$ in~(\ref{eq:trans-scalar}) has
disappeared altogether.  This implies that one can impose initial
values $\delta_\varepsilon(t_0,\boldsymbol{x})$,
$\dot{\delta}_\varepsilon(t_0,\boldsymbol{x})$ and $\delta_n(t_0,\boldsymbol{x})$
which can, in principle, be obtained from observation and,
subsequently, calculate the evolution of the true, physical, density
contrast functions $\delta_\varepsilon(t,\boldsymbol{x})$ and
$\delta_n(t,\boldsymbol{x})$.

The background equations~(\ref{subeq:einstein-flrw}) and the
perturbation equations~(\ref{subeq:final}) constitute a set of
equations which enables one to study the evolution of small
fluctuations in the energy density $\delta_\varepsilon$ and the
particle number density $\delta_n$ in an open, flat and closed
\textsc{flrw} universe with $\Lambda\not\equiv0$ and filled with a
perfect fluid described by an equation of state for the
pressure $p=p(n,\varepsilon)$.  This will be the subject of the
companion article \citep{2016arXiv160101260M}.


\section{Thermodynamics}
\label{sec:thermodyn}

In this section expressions for the true, physical pressure and
temperature perturbations are derived.  It is shown that density
perturbations are adiabatic if and only if the particle number density
does not contribute to the pressure.

\subsection{Pressure and Temperature Perturbations}
\label{sec:gi-PT}

From the equation of state for the pressure $p=p(n,\varepsilon)$, one
has
$\dot{p}_\nul=p_n\dot{n}_\nul+p_\varepsilon\dot{\varepsilon}_\nul$.
Multiplying both sides of this expression by
$\theta_\een/\dot{\theta}_\nul$ and subtracting the result from
$p_\een$ given by~(\ref{eq:p1}), one gets, using also~(\ref{e-n-gi}),
\begin{equation}
  \label{eq:def-pgi}
  p_\een-\dfrac{\dot{p}_\nul}{\dot{\theta}_\nul}\theta_\een=
      p_nn^\true_\een+p_\varepsilon\varepsilon^\true_\een.
\end{equation}
Hence, the quantity defined by
\begin{equation}
  \label{eq:gi-p}
  p^\true_\een\coloneqq p_\een-\dfrac{\dot{p}_\nul}{\dot{\theta}_\nul}\theta_\een,
\end{equation}
is the real pressure perturbation.  Combining~(\ref{eq:def-pgi})
and~(\ref{eq:gi-p}) and eliminating $p_\varepsilon$ with the help
of~(\ref{eq:beta-matter}), one arrives at
\begin{equation}
  \label{eq:p-dia-adia}
  p_\een^\true=\beta^2\varepsilon_\nul\delta_\varepsilon+n_\nul
  p_n\left[\delta_n-\dfrac{\delta_\varepsilon}{1+w}  \right],
\end{equation}
where also~(\ref{eq:contrast}) has been used.  The first term in this
expression is the \emph{adiabatic} part of the pressure perturbation
and the second term is the \emph{diabatic} part, as will be
discussed in the next two subsections.

From the equation of state~(\ref{eq:es-p-T}) for the energy density
$\varepsilon=\varepsilon(n,T)$ it follows that
\begin{align}
\label{eq:e1-dot-e0}
  \dot{\varepsilon}_\nul=\left(\dfrac{\partial\varepsilon}{\partial n}
  \right)_{\!\!T} \dot{n}_\nul+
   \left(\dfrac{\partial\varepsilon}{\partial T} \right)_{\!\!n}\dot{T}_\nul, \quad
  \varepsilon_\een=\left(\dfrac{\partial\varepsilon}{\partial n}
  \right)_{\!\!T}  n_\een+
   \left(\dfrac{\partial\varepsilon}{\partial T} \right)_{\!\!n} T_\een.
\end{align}
Multiplying $\dot{\varepsilon}_\nul$ by
$\theta_\een/\dot{\theta}_\nul$ and subtracting the result from
$\varepsilon_\een$, one finds, using~(\ref{eq:e-true-and-n-true}),
\begin{equation}
  \label{eq:time-e}
  \varepsilon^\true_\een=
     \left(\dfrac{\partial\varepsilon}{\partial n}\right)_{\!\!T} n^\true_\een+
\left(\dfrac{\partial\varepsilon}{\partial T}\right)_{\!\!n}
\left[T_\een-\dfrac{\dot{T}_\nul}{\dot{\theta}_\nul}\theta_\een \right],
\end{equation}
implying that the quantity defined by
\begin{equation}
  \label{eq:gi-T}
  T^\true_\een\coloneqq T_\een-\dfrac{\dot{T}_\nul}{\dot{\theta}_\nul}\theta_\een,
\end{equation}
is the real temperature perturbation.  The expressions~(\ref{eq:gi-p})
and~(\ref{eq:gi-T}) are both of the form~(\ref{eq:e-true-and-n-true}).

\subsection{Diabatic Density Perturbations}
\label{sec:diabatic}

The combined First and Second Law of Thermodynamics is given by
\begin{equation}
  \label{eq:combined-fs-thermo}
  {\text{d}}E=T{\text{d}}S-p{\text{d}}V+\mu{\text{d}}N,
\end{equation}
where $E$, $S$ and $N$ are the energy, the entropy and the number of
particles of a system with volume $V$ and pressure $p$, and where
$\mu$, the thermal---or chemical---potential, is the energy needed to
add one particle to the system.  In terms of the particle number
density $n=N/V$, the energy per particle $E/N=\varepsilon/n$ and the
entropy per particle $s=S/N$ the law~(\ref{eq:combined-fs-thermo}) can
be rewritten as
\begin{equation}
  \label{eq:law-rewritten}
  {\text{d}}\left(\dfrac{\varepsilon}{n}N\right)=T{\text{d}}(sN)-
    p{\text{d}}\left(\dfrac{N}{n}\right)+\mu{\text{d}}N,
\end{equation}
where $\varepsilon$ is the energy density.  The system is
\emph{extensive}, i.e.,
$S(\lambda E, \lambda V, \lambda N)=\lambda S(E,V,N)$, implying that
the entropy of the gas is $S=(E+pV-\mu N)/T$.  Dividing this relation
by $N$ one gets the so-called Euler relation
\begin{equation}
  \label{eq:chemical-pot}
  \mu=\dfrac{\varepsilon+p}{n}-Ts.
\end{equation}
Eliminating $\mu$ in~(\ref{eq:law-rewritten}) with the help
of~(\ref{eq:chemical-pot}), one finds that the combined First and
Second Law of Thermodynamics~(\ref{eq:combined-fs-thermo}) can be cast
in a form without $\mu$ and $N$, i.e.,
\begin{equation}
  \label{eq:thermo}
  T{\text{d}} s= {\text{d}}\Bigl(\dfrac{\varepsilon}{n}\Bigr)+
      p{\text{d}}\Bigl(\dfrac{1}{n}\Bigr).
\end{equation}
From the background equations~(\ref{subeq:einstein-flrw}) and the
thermodynamic law~(\ref{eq:thermo}) it follows that $\dot{s}_\nul=0$,
implying that the expansion of the universe takes place without
generating entropy.  Using~(\ref{eq:trans-scalar}) one finds that
$s_\een=s^{\text{gi}}_\een$ is automatically
gauge-invariant. Consequently, $s^{\text{gi}}_\een\equiv s_\een^\true$
is the true, physical perturbation to the entropy per particle. The
thermodynamic relation~(\ref{eq:thermo}) can,
using~(\ref{eq:contrast}) and $w\coloneqq p_\nul/\varepsilon_\nul$, be
rewritten as
\begin{equation}
  \label{eq:thermo-een}
   T_\nul s^\true_\een=-\dfrac{\varepsilon_\nul(1+w)}{n_\nul}
   \left[\delta_n-\dfrac{\delta_\varepsilon}{1+w}\right].
\end{equation}
From~(\ref{eq:thermo-een}) it follows that the right-hand side of
equation~(\ref{sec-ord}) is related to local perturbations in the
entropy, and equation~(\ref{fir-ord}) can be considered as an
evolution equation for entropy perturbations.  For
${T_\nul s^\true_\een\neq0}$ density perturbations are diabatic,
i.e., they exchange heat with their environment.

\subsection{Adiabatic Density Perturbations}
\label{subsec:adiabatic-pert}

In the limiting case that the contribution of the particle number
density to the pressure is negligible, i.e., $p_n\approx0$, the
coefficient $b_3$,~(\ref{eq:b3}), vanishes, so that the evolution
equation~(\ref{sec-ord}) becomes \emph{homogeneous}.  In this case
equation~(\ref{sec-ord}) describes a \emph{closed} system which does
not exchange heat with its environment.  In other words, the system is
\emph{adiabatic} and evolves only under its own gravity.

For $p_n\approx0$, the right-hand side of the entropy evolution
equation~(\ref{fir-ord}) vanishes.  Since ${T_\nul s^\true_\een=0}$
for adiabatic perturbations, one finds from~(\ref{eq:thermo-een}) that
the solution of~(\ref{fir-ord}) is
\begin{equation}
  \label{eq:adiabatic-condition}
  \delta_n(t,\boldsymbol{x})-\dfrac{\delta_\varepsilon(t,\boldsymbol{x})}{1+w(t)}=0.
\end{equation}
This relation implies that perturbations in the particle number
density, $\delta_n$, are through gravitation coupled to perturbations
in the (total) energy density, $\delta_\varepsilon$,
\emph{independent} of the nature of the particles.

The pressure perturbation~(\ref{eq:p-dia-adia}) becomes
$p_\een^\true=\beta^2\varepsilon_\nul\delta_\varepsilon$, i.e., only
the adiabatic part of the pressure perturbation survives.  Therefore,
in a fluid described by a barotropic equation of state,
$p\approx p(\varepsilon)$, density perturbations evolve adiabatically.
In all other cases where $p=p(n,\varepsilon)$, local density
perturbations evolve \emph{diabatically}.

\section{Conclusion}

\label{sec:conclusion}

In this article a new approach to the theory of cosmological
perturbations in \textsc{flrw} universes has been developed.  By using
the decomposition theorems, Section~\ref{sec:decomp-h-u}, to the
fullest extent, the set of linearized Einstein equations and
conservation laws can for scalar perturbations be recast in the
form~(\ref{subeq:pertub-flrw}) which is the perturbed counterpart of
the background Einstein equations and conservation
laws~(\ref{subeq:einstein-flrw}).  From the systems of
equations~(\ref{subeq:einstein-flrw}) and~(\ref{subeq:pertub-flrw}) it
follows that only three scalars $\varepsilon$, $n$ and $\theta$ and
their perturbations play a role in the evolution of density
perturbations.  Using the transformation rule~(\ref{eq:trans-scalar})
one can construct one and only one non-trivial set of gauge-invariant
quantities~(\ref{subeq:gi-quant}).  Since the General Theory of
Relativity is a mathematically consistent theory, there exists one and
only one quantity $\varepsilon^\true_\een$ for the perturbation to the
energy density and one and only one quantity $n^\true_\een$ for the
perturbation to the particle number density.  Therefore, the unique
gauge-invariant quantities $\varepsilon_\een^{\text{gi}}$ and
$n_\een^{\text{gi}}$ found in Section~\ref{sec:unique} must be equal
to the true physical density perturbations $\varepsilon^\true_\een$
and $n^\true_\een$, respectively.  In Section~\ref{sec:newt-limit} on the
non-relativistic limit it is shown that this is indeed the case. In
fact, it is shown that the system of
equations~(\ref{subeq:pertub-flrw}) and the definitions~(\ref{e-n-gi})
reduce in the non-relativistic to the time-independent Poisson
equation~(\ref{eq:poisson}) and the relation~(\ref{eq:newt-ngi}).  The
gauge transformation of the General Theory of Relativity reduces in
the non-relativistic limit to the gauge transformation of the
Newtonian Theory of Gravity.  That is why in the General Theory of
Relativity a coordinate system cannot be uniquely fixed, since the
Newtonian Theory of Gravity is invariant under the residual gauge
transformation~(\ref{eq:gauge-trans-newt}).

In Section~\ref{sec:flat-pert} evolution equations for density
perturbations $\varepsilon_\een^\true$ and $n_\een^\true$ in closed,
flat and open \textsc{flrw} universes are derived.  What is done,
essentially, is to rewrite the linearized Einstein equations and
conservation laws for the quantities $\varepsilon_\een$ and $n_\een$
in terms of the measurable energy and particle number densities
$\varepsilon_\een^\true$ and $n_\een^\true$, just as in
electromagnetism the Coulomb and Ampère laws for the scalar
potential~$\Phi$ and vector potential $\boldsymbol{A}$ can be rewritten in the
form of the Maxwell equations for the measurable electric field
$\boldsymbol{E}$ and magnetic field $\boldsymbol{B}$.

The treatment of cosmological perturbations presented in this article
has the following properties.  Firstly, in
equations~(\ref{subeq:pertub-flrw}) \emph{all metric tensor
  perturbations} are contained in only three quantities:
$R_\een$,~(\ref{driekrom}), $\theta_\een$ and
$\vartheta_\een$,~(\ref{fes5}).  This eliminates completely the
unwanted gradient terms \citep{2013GReGr..45.1989M} in the resulting
equations~(\ref{subeq:final}).  Furthermore, the
quantities~(\ref{eq:e-true-and-n-true}) do not contain spatial
derivatives \citep{Ellis1, Ellis2, ellis-1998}. Secondly, due to the
quotients $\dot{\varepsilon}_\nul/\dot{\theta}_\nul$ and
$\dot{n}_\nul/\dot{\theta}_\nul$ of time derivatives, the quantities
$\varepsilon_\een^\true$ and
$n_\een^\true$,~(\ref{eq:e-true-and-n-true}), are independent of the
definition of time, so that their evolution is only determined by
their propagation equations.  Finally, from~(\ref{eq:theta-phys}) it
follows that, in the linear regime, the \emph{global} expansion
$\theta_\nul=3H$ is not affected by a \emph{local} density
perturbation. This is in accordance with the results found
by~\cite{2014arXiv1407.8084G}.

In a fluid described by a barotropic equation of state,
$p\approx p(\varepsilon)$, the right-hand side of the evolution
equation~(\ref{sec-ord}) vanishes and equation~(\ref{fir-ord}) reduces
to an identity, so that density perturbations evolve adiabatically.
In all other cases where $p=p(n,\varepsilon)$, local density
perturbations evolve diabatically.

In this article an equation of state for the pressure
$p(n,\varepsilon)$ in a perfect fluid is used. However, the algorithm
in the appendix can be easily adapted to more general cases.

The equations~(\ref{subeq:final}) have been checked using the computer
algebra system~\cite{maxima}, as follows.  Substituting the contrast
functions~(\ref{eq:contrast}) into equations~(\ref{subeq:final}),
where $\varepsilon_\een^\true$ and $n_\een^\true$ are given
by~(\ref{subeq:pertub-gi-e-n}), and subsequently eliminating the
time-derivatives of $\varepsilon_\nul$, $n_\nul$, $H$, $R_\nul$ and
$\varepsilon_\een$, $n_\een$, $\vartheta_\een$, $R_\een$ with the help
of equations~(\ref{subeq:einstein-flrw}) and~(\ref{subeq:pertub-gi}),
respectively, yields two identities for each of the two
equations~(\ref{subeq:final}).

In a companion article~\citep{2016arXiv160101260M} the evolution
equations~(\ref{subeq:final}) will be solved.  Since the gauge problem
of cosmology is solved, the physical consequences of the new approach
to cosmological perturbations stand out clearly.

\appendix
\numberwithin{equation}{section}

\section{Derivation of the Evolution Equations using Computer Algebra}

In this appendix the perturbation equations~(\ref{subeq:final}) of the
main text are derived from the basic perturbation
equations~(\ref{subeq:pertub-gi}) and the
definitions~(\ref{eq:contrast}).  This is done by first deriving the
evolution equations for the gauge-invariant quantities
$\varepsilon^\true_\een$ and $n^\true_\een$~(\ref{subeq:pertub-gi-e-n}):
\begin{subequations}
\label{subeq:eerste}
\begin{align}
  \ddot{\varepsilon}^\true_\een+a_1\dot{\varepsilon}^\true_\een+
  a_2\varepsilon^\true_\een & = a_3 \left(n_\een^\true -
  \frac{n_\nul}{\varepsilon_\nul(1+w)}\varepsilon_\een^\true\right),
      \label{eq:vondst2}  \\
  \dfrac{1}{c}\dfrac{\text{d}}{\text{d}t}\left(n_\een^\true -
  \frac{n_\nul}{\varepsilon_\nul(1+w)}\varepsilon_\een^\true\right)&=
    -3H\left(1-\frac{n_\nul
  p_n}{\varepsilon_\nul(1+w)}\right) \left(n_\een^\true -
  \frac{n_\nul}{\varepsilon_\nul(1+w)}\varepsilon_\een^\true\right).   \label{eq:vondst1}
\end{align}
\end{subequations}
The coefficients $a_1$, $a_2$ and $a_3$ occurring in
equation~(\ref{eq:vondst2}) are given~by
\begin{subequations}
\label{subeq:coeff}
\begin{align}
  a_1 & = \dfrac{\kappa\varepsilon_\nul(1+w)}{H}
  -2\dfrac{\dot{\beta}}{\beta}+H(4-3\beta^2)
    +R_\nul\left(\dfrac{1}{3H} + \dfrac{2H(1+3\beta^2)}
  {R_\nul+3\kappa\varepsilon_\nul(1+w)}\right), \label{eq:alpha-1} \\
  a_2 & = \kappa\varepsilon_\nul(1+w)-
  4H\dfrac{\dot{\beta}}{\beta}+2H^2(2-3\beta^2) 
   +R_\nul\left(\dfrac{1}{2}+
\dfrac{5H^2(1+3\beta^2)-2H\dfrac{\dot{\beta}}{\beta}}
  {R_\nul+3\kappa\varepsilon_\nul(1+w)}\right)
-\beta^2\left(\frac{\tilde{\nabla}^2}{a^2}-\tfrac{1}{2}R_\nul \right), \label{eq:alpha-2} \\
  a_3 & =\,
\Biggl\{\dfrac{-18H^2}{R_\nul+3\kappa\varepsilon_\nul(1+w)}
  \Biggl[\varepsilon_\nul p_{\varepsilon n}(1+w) 
   +\dfrac{2p_n}{3H}\dfrac{\dot{\beta}}{\beta}
    +p_n(p_\varepsilon-\beta^2)+n_\nul p_{nn}\Biggr]+
   p_n\Biggr\}
\left(\frac{\tilde{\nabla}^2}{a^2}-\tfrac{1}{2}R_\nul\right).
       \label{eq:alpha-3}
\end{align}
\end{subequations}
In calculating the coefficients $a_1$, $a_2$
and~$a_3$,~(\ref{subeq:coeff}), it is used that the time derivative of
the quotient $w\coloneqq p_\nul/\varepsilon_\nul$ is given
by~(\ref{eq:time-w}).  Moreover, it is convenient \emph{not} to expand
the quantity $\beta^2\coloneqq\dot{p}_\nul/\dot{\varepsilon}_\nul$
since this considerably complicates the expressions for the
coefficients~$a_1$, $a_2$ and~$a_3$.

\subsection{Derivation of the Evolution Equation for the Energy
  Density Perturbation}

In order to derive equation~(\ref{eq:vondst2}), the
system~(\ref{subeq:pertub-gi}) and expression~(\ref{Egi}) is
rewritten, using~(\ref{eq:p1}), in the form
\begin{subequations}
\label{subeq:nieuw}
\begin{align}
  \dot{\varepsilon}_\een+\alpha_{11}\varepsilon_\een+
    \alpha_{12}n_\een+
   \alpha_{13}\vartheta_\een+\alpha_{14}R_\een & = 0,
\label{nieuw1} \\
   \dot{n}_\een+\alpha_{21}\varepsilon_\een +
    \alpha_{22}n_\een+
   \alpha_{23}\vartheta_\een+\alpha_{24}R_\een & = 0,
\label{nieuw2} \\
  \dot{\vartheta}_\een +\alpha_{31}\varepsilon_\een +
     \alpha_{32}n_\een+
   \alpha_{33}\vartheta_\een+\alpha_{34}R_\een & = 0,
\label{nieuw3} \\
    \dot{R}_\een+\alpha_{41}\varepsilon_\een+
     \alpha_{42}n_\een+
   \alpha_{43}\vartheta_\een+\alpha_{44}R_\een & = 0,
\label{nieuw4} \\
  \varepsilon^\true_\een+\alpha_{51}\varepsilon_\een+\alpha_{52}n_\een+
    \alpha_{53}\vartheta_\een+\alpha_{54}R_\een & = 0,
\label{nieuw5}
\end{align}
\end{subequations}
where the coefficients $\alpha_{ij}$ are given in Table~\ref{eq:aij}.

\begin{table*}
\renewcommand{\arraystretch}{1} 
\normalsize   
\caption{\label{eq:aij}The coefficients $\alpha_{ij}$ figuring in
the equations~(\ref{subeq:nieuw}).}
\[  \begin{array}{cccc} \hline\hline
  & & & \\
  3H(1+p_\varepsilon)+\dfrac{\kappa\varepsilon_\nul(1+w)}{2H} &
      3Hp_n & \varepsilon_\nul(1+w) &
      \dfrac{\varepsilon_\nul(1+w)}{4H} \\  
 & & & \\
  \dfrac{\kappa n_\nul}{2H} & 3H & n_\nul & \dfrac{n_\nul}{4H} \\  
 & & & \\
  \dfrac{p_\varepsilon}{\varepsilon_\nul(1+w)}\dfrac{\tilde{\nabla}^2}{a^2} &
       \dfrac{p_n}{\varepsilon_\nul(1+w)}\dfrac{\tilde{\nabla}^2}{a^2} &
       H(2-3\beta^2) & 0 \\
 & & & \\
  \dfrac{\kappa R_\nul}{3H} & 0 &
  -2\kappa\varepsilon_\nul(1+w) & 2H+\dfrac{R_\nul}{6H} \\
 & & & \\
  \dfrac{-R_\nul}{R_\nul+3\kappa\varepsilon_\nul(1+w)} & 0 &
     \dfrac{6\varepsilon_\nul H(1+w)}{R_\nul+3\kappa\varepsilon_\nul(1+w)} &
     \dfrac{\tfrac{3}{2}\varepsilon_\nul(1+w)}{R_\nul+3\kappa\varepsilon_\nul(1+w)}  \\
 & & & \\  \hline\hline
\end{array}  \]
\end{table*}

\paragraph{Step 1} First the quantity $R_\een$ is eliminated from
equations~(\ref{subeq:nieuw}). Differentiating equation~(\ref{nieuw5})
with respect to time and eliminating the time derivatives
$\dot{\varepsilon}_\een$, $\dot{n}_\een$, $\dot{\vartheta}_\een$
and~$\dot{R}_\een$ with the help of equations
(\ref{nieuw1})--(\ref{nieuw4}), one arrives at the equation
\begin{equation}\label{eq:equiv}
   \dot{\varepsilon}^\true_\een + p_1\varepsilon_\een+p_2 n_\een+
   p_3\vartheta_\een+p_4 R_\een=0,
\end{equation}
where the coefficients $p_1,\ldots,p_4$ are given by
\begin{equation}\label{eq:coef-pi}
  p_i=\dot{\alpha}_{5i}-\alpha_{51}\alpha_{1i}-
         \alpha_{52}\alpha_{2i}-\alpha_{53}\alpha_{3i}-\alpha_{54}\alpha_{4i}.
\end{equation}
From equation~(\ref{eq:equiv}) it follows that
\begin{equation}\label{eq:sol-3R1}
   R_\een=-\dfrac{1}{p_4}\dot{\varepsilon}^\true_\een-
     \dfrac{p_1}{p_4}\varepsilon_\een-\dfrac{p_2}{p_4}n_\een-
     \dfrac{p_3}{p_4}\vartheta_\een.
\end{equation}
In this way the quantity $R_\een$ has been expressed as a linear
combination of the quantities $\dot{\varepsilon}^\true_\een$,
$\varepsilon_\een$, $n_\een$ and $\vartheta_\een$. Upon replacing
$R_\een$ in equations~(\ref{subeq:nieuw}) by the right-hand side
of~(\ref{eq:sol-3R1}), one arrives at the system of equations
\begin{subequations}
\label{subeq:tweede}
\begin{align}
 \dot{\varepsilon}_\een+q_1\dot{\varepsilon}_\een^\true+
   \gamma_{11}\varepsilon_\een+\gamma_{12}n_\een+
   \gamma_{13}\vartheta_\een & = 0, \label{tweede1} \\
 \dot{n}_\een+q_2\dot{\varepsilon}_\een^\true+
    \gamma_{21}\varepsilon_\een+\gamma_{22}n_\een+
    \gamma_{23}\vartheta_\een & = 0, \label{tweede2} \\
 \dot{\vartheta}_\een+q_3\dot{\varepsilon}_\een^\true+
   \gamma_{31}\varepsilon_\een+\gamma_{32}n_\een+
   \gamma_{33}\vartheta_\een & = 0, \label{tweede3} \\
     \dot{R}_\een+
   q_4\dot{\varepsilon}^\true_\een+\gamma_{41}\varepsilon_\een+\gamma_{42}n_\een+
   \gamma_{43}\vartheta_\een & = 0, \label{tweede4} \\
 \varepsilon^\true_\een+
   q_5\dot{\varepsilon}^\true_\een+\gamma_{51}\varepsilon_\een+\gamma_{52}n_\een+
   \gamma_{53}\vartheta_\een & = 0, \label{tweede5}
\end{align}
\end{subequations}
where the coefficients $q_i$ and $\gamma_{ij}$ are given by
\begin{equation}\label{eq:betaij}
  q_i=-\dfrac{\alpha_{i4}}{p_4}, \quad
   \gamma_{ij}=\alpha_{ij}+q_i p_j.
\end{equation}
It has now been achieved that the quantity $R_\een$ occurs
\emph{explicitly} only in equation~(\ref{tweede4}), whereas $R_\een$
occurs \emph{implicitly} in the remaining equations. Therefore,
equation~(\ref{tweede4}) is not needed anymore.  Equations
(\ref{tweede1})--(\ref{tweede3}) and~(\ref{tweede5}) are four ordinary
differential equations for the four unknown quantities
$\varepsilon_\een$, $n_\een$, $\vartheta_\een$ and
$\varepsilon^\true_\een$.

\paragraph{Step 2} In the same way as in Step~1, the explicit
occurrence of the quantity $\vartheta_\een$ is eliminated from the
system of equations~(\ref{subeq:tweede}).  Differentiating
equation~(\ref{tweede5}) with respect to time and eliminating the time
derivatives $\dot{\varepsilon}_\een$, $\dot{n}_\een$ and
$\dot{\vartheta}_\een$ with the help of equations
(\ref{tweede1})--(\ref{tweede3}), one arrives at
\begin{equation}
\label{eq:ddot-egi}
  q_5\ddot{\varepsilon}^\true_\een+r\dot{\varepsilon}^\true_\een+
     s_1\varepsilon_\een+s_2n_\een+s_3\vartheta_\een=0,
\end{equation}
where the coefficients $s_i$ and $r$ are given by
\begin{subequations}
\label{eq:coef-qi}
\begin{align}
  s_i & = \dot{\gamma}_{5i}-\gamma_{51}\gamma_{1i}-\gamma_{52}\gamma_{2i}-
       \gamma_{53}\gamma_{3i}, \\
  r & = 1+\dot{q}_5-\gamma_{51}q_1-\gamma_{52}q_2-\gamma_{53}q_3.
\end{align}
\end{subequations}
From equation~(\ref{eq:ddot-egi}) it follows that
\begin{equation}\label{eq:sol-theta1}
  \vartheta\een=-\dfrac{q_5}{s_3}\ddot{\varepsilon}^\true_\een-
     \dfrac{r}{s_3}\dot{\varepsilon}^\true_\een-
     \dfrac{s_1}{s_3}\varepsilon_\een-\dfrac{s_2}{s_3}n_\een.
\end{equation}
In this way the quantity $\vartheta_\een$ is expressed as a linear
combination of the quantities $\ddot{\varepsilon}^\true_\een$,
$\dot{\varepsilon}^\true_\een$, $\varepsilon_\een$ and $n_\een$. Upon
replacing $\vartheta_\een$ in equations~(\ref{subeq:tweede}) by the
right-hand side of~(\ref{eq:sol-theta1}), one arrives at the system of
equations
\begin{subequations}
\label{subeq:derde}
\begin{align}
\dot{\varepsilon}_\een-\gamma_{13}\dfrac{q_5}{s_3}\ddot{\varepsilon}^\true_\een+
   \left(q_1-\gamma_{13}\dfrac{r}{s_3}\right)\dot{\varepsilon}^\true_\een
  +\left(\gamma_{11}-\gamma_{13}\dfrac{s_1}{s_3}\right)\varepsilon_\een
  +\left(\gamma_{12}-\gamma_{13}\dfrac{s_2}{s_3}\right)n_\een & =0, \label{derde1}
\\
\dot{n}_\een-\gamma_{23}\dfrac{q_5}{s_3}\ddot{\varepsilon}^\true_\een+
   \left(q_2-\gamma_{23}\dfrac{r}{s_3}\right)\dot{\varepsilon}^\true_\een
    +\left(\gamma_{21}-\gamma_{23}\dfrac{s_1}{s_3}\right)\varepsilon_\een
  +\left(\gamma_{22}-\gamma_{23}\dfrac{s_2}{s_3}\right)n_\een & =0, \label{derde2}
\\
\dot{\vartheta}_\een-\gamma_{33}\dfrac{q_5}{s_3}\ddot{\varepsilon}^\true_\een+
   \left(q_3-\gamma_{33}\dfrac{r}{s_3}\right)\dot{\varepsilon}^\true_\een
   +\left(\gamma_{31}-\gamma_{33}\dfrac{s_1}{s_3}\right)\varepsilon_\een
  +\left(\gamma_{32}-\gamma_{33}\dfrac{s_2}{s_3}\right)n_\een & =0, \label{derde3}
\\
  \dot{R}_\een-\gamma_{43}\dfrac{q_5}{s_3}\ddot{\varepsilon}^\true_\een+
   \left(q_4-\gamma_{43}\dfrac{r}{s_3}\right)\dot{\varepsilon}^\true_\een
  + \left(\gamma_{41}-\gamma_{43}\dfrac{s_1}{s_3}\right)\varepsilon_\een
  +\left(\gamma_{42}-\gamma_{43}\dfrac{s_2}{s_3}\right)n_\een & =0, \label{derde4}
\\
\varepsilon^\true_\een-\gamma_{53}\dfrac{q_5}{s_3}\ddot{\varepsilon}^\true_\een+
   \left(q_5-\gamma_{53}\dfrac{r}{s_3}\right)\dot{\varepsilon}^\true_\een
    +\left(\gamma_{51}-\gamma_{53}\dfrac{s_1}{s_3}\right)\varepsilon_\een
  +\left(\gamma_{52}-\gamma_{53}\dfrac{s_2}{s_3}\right)n_\een & =0. \label{derde5}
\end{align}
\end{subequations}
It has now been achieved that the quantities $\vartheta_\een$ and
$R_\een$ occur \emph{explicitly} only in equations~(\ref{derde3})
and~(\ref{derde4}), whereas they occur \emph{implicitly} in the
remaining equations.  Therefore, equations~(\ref{derde3})
and~(\ref{derde4}) are not needed anymore.  Equations~(\ref{derde1}),
(\ref{derde2}) and~(\ref{derde5}) are three ordinary differential
equations for the three unknown quantities $\varepsilon_\een$,
$n_\een$ and $\varepsilon^\true_\een$.

\paragraph{Step 3} At first sight, the next steps would be to
eliminate, successively, the quantities $\varepsilon_\een$ and
$n_\een$ from equation~(\ref{derde5}) with the help of
equations~(\ref{derde1}) and~(\ref{derde2}).  One would then end up
with a fourth-order differential equation for the unknown quantity
$\varepsilon^\true_\een$.  This, however, is impossible, since the
gauge-dependent quantities $\varepsilon_\een$ and $n_\een$ do
\emph{not} occur explicitly in equation~(\ref{derde5}), as is now
shown.  Firstly, it is observed that equation~(\ref{derde5}) can be
rewritten as
\begin{equation}\label{eq:eindelijk}
  \ddot{\varepsilon}^\true_\een+a_1\dot{\varepsilon}^\true_\een+
    a_2\varepsilon^\true_\een=
    a_3\left(n_\een+\dfrac{\gamma_{51}s_3-\gamma_{53}s_1}
    {\gamma_{52}s_3-\gamma_{53}s_2} \varepsilon_\een\right),
\end{equation}
where the coefficients $a_1$, $a_2$ and $a_3$ are given by
\begin{equation}
\label{subeq:vierde}
  a_1  = -\dfrac{s_3}{\gamma_{53}}+\dfrac{r}{q_5}, \quad
  a_2  = -\dfrac{s_3}{\gamma_{53}q_5}, \quad
  a_3  = \dfrac{\gamma_{52}s_3}{\gamma_{53}q_5}-\dfrac{s_2}{q_5}.
\end{equation}
These are the coefficients~(\ref{subeq:coeff}).  Secondly, one finds
\begin{equation}\label{eq:check}
  \dfrac{\gamma_{51}s_3-\gamma_{53}s_1}{\gamma_{52}s_3-\gamma_{53}s_2}=
      -\dfrac{n_\nul}{\varepsilon_\nul(1+w)}.
\end{equation}
Finally, using the expressions~(\ref{eq:e-true-and-n-true}) and the conservation
laws~(\ref{FRW2}) and~(\ref{FRW2a}), one gets
\begin{equation}
  \label{eq:equal-gi-non-gi}
  n_\een-\dfrac{n_\nul}{\varepsilon_\nul(1+w)}\varepsilon_\een=
  n^\true_\een-\dfrac{n_\nul}{\varepsilon_\nul(1+w)}\varepsilon^\true_\een.
\end{equation}
Consequently, the right-hand side of~(\ref{eq:eindelijk}) does not
explicitly contain the gauge-dependent quantities $\varepsilon_\een$
and $n_\een$.  With the help of expression~(\ref{eq:equal-gi-non-gi})
one can rewrite equation~(\ref{eq:eindelijk}) in the
form~(\ref{eq:vondst2}).

The derivation of the coefficients~(\ref{subeq:coeff})
from~(\ref{subeq:vierde}) and the proof of the
equality~(\ref{eq:check}) is straightforward, but extremely
complicated.  The computer algebra system
\cite{maxima} has been used to perform
this algebraic task.

\subsection{Derivation of the Evolution Equation for the Entropy
  Perturbation}

The basic set of equations~(\ref{subeq:pertub-gi}) from which the
evolution equations are derived is of fourth-order.  From this system
a second-order equation~(\ref{eq:vondst2}) for
$\varepsilon^\true_\een$ has been extracted.  Therefore, the remaining
system from which an evolution equation for $n^\true_\een$ can be
derived is at most of second order.  Since gauge-invariant quantities
$\varepsilon^\true_\een$ and $n^\true_\een$ have been used, one degree
of freedom, namely the gauge function $\xi^0(t,\boldsymbol{x})$
in~(\ref{eq:trans-scalar}) has disappeared.  As a consequence, only a
first-order evolution equation for $n^\true_\een$ can be derived.
Instead of deriving an equation for $n^\true_\een$, an evolution
equation~(\ref{eq:vondst1}) for the entropy perturbation, which
contains $n^\true_\een$, is derived.

From the combined First and Second Law of
Thermodynamics~(\ref{eq:thermo}) it follows that
\begin{equation}
  \label{eq:thermo-n0-e0}
  T_\nul s_\een=-\dfrac{\varepsilon_\nul(1+w)}{n^2_\nul}\left[n_\een-
       \dfrac{n_\nul}{\varepsilon_\nul(1+w)}\varepsilon_\een\right],
\end{equation}
where the right-hand side is gauge-invariant by virtue
of~(\ref{eq:equal-gi-non-gi}), so that $s_\een=s^\true_\een$ is
gauge-invariant, in accordance with the remark
below~(\ref{eq:thermo}).  Differentiating the term between square
brackets in~(\ref{eq:thermo-n0-e0}) with respect to time and using the
background equations~(\ref{FRW2}) and~(\ref{FRW2a}), the first-order
equations~(\ref{FRW4gi}) and~(\ref{FRW4agi}) and the definitions
$w\coloneqq p_\nul/\varepsilon_\nul$ and
$\beta^2\coloneqq\dot{p}_\nul/\dot{\varepsilon}_\nul$, one finds
\begin{align}
  \label{eq:vondst-gauge}
    \dfrac{1}{c}\dfrac{\text{d}}{\text{d}t}\left(n_\een -
  \frac{n_\nul}{\varepsilon_\nul(1+w)}\varepsilon_\een\right)=
    -3H\left(1-\frac{n_\nul
  p_n}{\varepsilon_\nul(1+w)}\right) \left(n_\een -
  \frac{n_\nul}{\varepsilon_\nul(1+w)}\varepsilon_\een\right),
\end{align}
where \cite{maxima} has been used to perform the algebraic task.  By
virtue of~(\ref{eq:equal-gi-non-gi}), one may in this equation replace
$n_\een$ and $\varepsilon_\een$ by $n^\true_\een$ and
$\varepsilon^\true_\een$, respectively, thus obtaining
equation~(\ref{eq:vondst1}).

\subsection{Evolution Equations for the Contrast Functions}
\label{app:contrast}

First the entropy equation~(\ref{fir-ord}) is derived. From the
definitions~(\ref{eq:contrast}) it follows that
\begin{equation}\label{eq:sgi-contrast}
  n_\een^\true-\frac{n_\nul}{\varepsilon_\nul(1+w)}\varepsilon_\een^\true=
       n_\nul\left(\delta_n-\dfrac{\delta_\varepsilon}{1+w}\right).
\end{equation}
Substituting this expression into equation~(\ref{eq:vondst-gauge})
one finds, using also~(\ref{eq:equal-gi-non-gi}),

\begin{align}\label{eq:diff-sgi}
  \dfrac{1}{c}\dfrac{\text{d}}{\text{d}t}
   \left[n_\nul\left(\delta_n-\dfrac{\delta_\varepsilon}{1+w} \right)\right]=
-3H\left(1-\frac{n_\nul p_n}{\varepsilon_\nul(1+w)}\right)
\left[n_\nul\left(\delta_n-\dfrac{\delta_\varepsilon}{1+w} \right)\right].
\end{align}
Using equation~(\ref{FRW2a}) one arrives at equation~(\ref{fir-ord})
of the main text.

Finally, equation~(\ref{sec-ord}) is derived. Upon substituting the
expressions
\begin{equation}
\label{subeq:afgeleiden}
\varepsilon^\true_\een  = \varepsilon_\nul\delta_\varepsilon,\quad
\dot{\varepsilon}^\true_\een=\dot{\varepsilon}_\nul\delta_\varepsilon+
\varepsilon_\nul\dot{\delta}_\varepsilon, \quad
      \ddot{\varepsilon}^\true_\een=\ddot{\varepsilon}_\nul\delta_\varepsilon+
      2\dot{\varepsilon}_\nul\dot{\delta}_\varepsilon+\varepsilon_\nul\ddot{\delta}_\varepsilon,
\end{equation}
into equation~(\ref{eq:vondst2}), and dividing by $\varepsilon_\nul$,
one finds
\begin{align}
 b_1=2\dfrac{\dot{\varepsilon}_\nul}{\varepsilon_\nul}+a_1, \quad
 b_2=\dfrac{\ddot{\varepsilon}_\nul}{\varepsilon_\nul}+
      a_1\dfrac{\dot{\varepsilon}_\nul}{\varepsilon_\nul}+a_2, \quad
 b_3=a_3\dfrac{n_\nul}{\varepsilon_\nul},
\end{align}
where also~(\ref{eq:sgi-contrast}) has been used. Using~\cite{maxima},
one arrives at the coefficients~(\ref{subeq:coeff-contrast}) of the
main text.


\end{document}